\documentclass[preprint2,twocolappendix]{aastex631}
\usepackage{CJK}
\DeclareTextSymbolDefault{\dh}{T1}
\usepackage{fontawesome}
\shortauthors{Ca\~nas et al.}
\shorttitle{TOI-3984 A b and TOI-5293 A b transit M dwarfs in wide binaries}
\graphicspath{{./}{figures/}}

\newcommand{\PSUAA}{Department of Astronomy \& Astrophysics, The Pennsylvania State University, 525 Davey Laboratory, University Park, PA 16802, USA}
\newcommand{\PSUCEHW}{Center for Exoplanets and Habitable Worlds, The Pennsylvania State University, 525 Davey Laboratory, University Park, PA 16802, USA}
\newcommand{\Goddard}{NASA Goddard Space Flight Center, 8800 Greenbelt Road, Greenbelt, MD 20771, USA}
\newcommand{\UWY}{Department of Physics \& Astronomy, University of Wyoming, Laramie, WY 82070, USA}
\newcommand{\STSCI}{Space Telescope Science Institute, 3700 San Martin Drive, Baltimore, MD 21218, USA}
\newcommand{\JH}{Department of Physics and Astronomy, Johns Hopkins University, 3400 N Charles St, Baltimore, MD 21218, USA}
\newcommand{\ETH}{ETH Zurich, Institute for Particle Physics \& Astrophysics, Zurich, Switzerland}
\newcommand{\UA}{Steward Observatory, The University of Arizona, 933 N.\ Cherry Avenue, Tucson, AZ 85721, USA}
\newcommand{\UC}{Department of Physics, 390 UCB, University of Colorado, Boulder, CO 80309, USA}
\newcommand{\NIST}{National Institute of Standards \& Technology, 325 Broadway, Boulder, CO 80305, USA}
\newcommand{\UCI}{Department of Physics \& Astronomy, The University of California, Irvine, Irvine, CA 92697, USA}
\newcommand{\Princeton}{Department of Astrophysical Sciences, Princeton University, 4 Ivy Lane, Princeton, NJ 08540, USA}
\newcommand{\UT}{McDonald Observatory and Center for Planetary Systems Habitability, The University of Texas at Austin, Austin, TX 78730, USA}

\begin{document}
\begin{CJK*}{UTF8}{gbsn}

\title{TOI-3984 A b and TOI-5293 A b: two temperate gas giants transiting mid-M dwarfs in wide binary systems}
\correspondingauthor{Caleb I. Ca\~nas}
\email{c.canas@nasa.gov}

\author[0000-0003-4835-0619]{Caleb I. Ca\~nas}
\altaffiliation{NASA Postdoctoral Fellow}
\affiliation{\Goddard}
\affiliation{\PSUAA}
\affiliation{\PSUCEHW}

\author[0000-0001-8401-4300]{Shubham Kanodia}
\affil{Earth and Planets Laboratory, Carnegie Institution for Science, 5241 Broad Branch Road, NW, Washington, DC 20015, USA}
\affiliation{\PSUAA}
\affiliation{\PSUCEHW}

\author[0000-0002-2990-7613]{Jessica Libby-Roberts}
\affiliation{\PSUAA}
\affiliation{\PSUCEHW}

\author[0000-0002-9082-6337]{Andrea S.J. Lin}
\affil{\PSUAA}
\affil{\PSUCEHW}

\author[0000-0003-2435-130X]{Maria Schutte}
\affil{Homer L. Dodge Department of Physics and Astronomy, University of Oklahoma, 440 W. Brooks Street, Norman, OK 73019, USA}

\author[0000-0002-5300-5353]{Luke Powers}
\affil{\PSUAA}
\affil{\PSUCEHW}

\author[0000-0002-7227-2334]{Sinclaire Jones}
\affil{Department of Astronomy, The Ohio State University, 4055 McPherson Laboratory, Columbus, OH 43210, USA}
\affil{\Princeton}

\author[0000-0002-0048-2586]{Andrew Monson}
\affil{\UA}

\author[0000-0002-7846-6981]{Songhu Wang}
\affiliation{Department of Astronomy, Indiana University, 727 East 3rd Street, Bloomington, IN 47405-7105, USA}

\author[0000-0001-7409-5688]{Gu\dh mundur Stef\'ansson}
\altaffiliation{NASA Sagan Fellow}
\affil{\Princeton}

\author[0000-0001-9662-3496]{William D. Cochran}
\affil{\UT}

\author[0000-0003-0149-9678]{Paul Robertson}
\affil{\UCI}

\author[0000-0001-9596-7983]{Suvrath Mahadevan}
\affil{\PSUAA}
\affil{\PSUCEHW}
\affil{\ETH}  

\author[0000-0001-7458-1176]{Adam F. Kowalski}
\affil{National Solar Observatory, University of Colorado Boulder, 3665 Discovery Drive, Boulder, CO 80303, USA}
\affil{Department of Astrophysical and Planetary Sciences, University of Colorado, Boulder, 2000 Colorado Ave, CO 80305, USA}
\affil{Laboratory for Atmospheric and Space Physics, University of Colorado Boulder, 3665 Discovery Drive, Boulder, CO 80303, USA}

\author[0000-0001-9209-1808]{John Wisniewski}
\affiliation{Department of Physics and Astronomy, George Mason University, Fairfax, VA 22030, USA}

\author[0000-0001-9307-8170]{Brock A. Parker}
\affil{\UWY}

\author[0000-0002-2401-8411]{Alexander Larsen}
\affil{\UWY}

\author{Franklin A.L. Chapman}
\affil{\UWY}

\author[0000-0002-4475-4176]{Henry A. Kobulnicky}
\affil{\UWY}

\author[0000-0002-5463-9980]{Arvind F. Gupta}
\affil{\PSUAA}
\affil{\PSUCEHW}

\author[0000-0002-0885-7215]{Mark E. Everett}
\affil{NSF's National Optical-Infrared Astronomy Research Laboratory, 950 N. Cherry Ave., Tucson, AZ 85719, USA}

\author[0000-0002-7000-3181]{Bryan Edward Penprase}
\affiliation{Soka University of America, Aliso Viejo CA, 92656, USA}

\author[0000-0003-2307-0629]{Gregory Zeimann}
\affil{Hobby Eberly Telescope, University of Texas at Austin, Austin, TX, 78712, USA}

\author[0000-0001-7708-2364]{Corey Beard}
\affiliation{\UCI}

\author[0000-0003-4384-7220]{Chad F.\ Bender}
\affil{\UA}

\author[0000-0001-8020-7121]{Knicole D. Col\'on}
\affil{\Goddard}

\author[0000-0002-5463-9980]{Scott A. Diddams}
\affil{Electrical, Computer \& Energy Engineering, 425 UCB, University of Colorado, Boulder, CO 80309, USA}
\affil{\UC}
\affil{\NIST}

\author[0000-0002-0560-1433]{Connor Fredrick}
\affil{\NIST}
\affil{\UC}

\author[0000-0003-1312-9391]{Samuel Halverson}
\affil{Jet Propulsion Laboratory, California Institute of Technology, 4800 Oak Grove Drive, Pasadena, California 91109}

\author[0000-0001-8720-5612]{Joe P.\ Ninan}
\affil{Department of Astronomy and Astrophysics, Tata Institute of Fundamental Research, Homi Bhabha Road, Colaba, Mumbai 400005, India}

\author[0000-0002-4289-7958]{Lawrence W. Ramsey}
\affil{\PSUAA}
\affil{\PSUCEHW}

\author[0000-0001-8127-5775]{Arpita Roy}
\affil{\STSCI}
\affil{\JH}

\author[0000-0002-4046-987X]{Christian Schwab}
\affil{School of Mathematical and Physical Sciences, Macquarie University, Balaclava Road, North Ryde, NSW 2109, Australia}



\begin{abstract}
We confirm the planetary nature of two gas giants discovered by TESS to transit M dwarfs with stellar companions at wide separations. TOI-3984 A ($J=11.93$) is an M4 dwarf hosting a short-period ($4.353326 \pm 0.000005$ days) gas giant ($M_p=0.14\pm0.03~\mathrm{M_{J}}$ and $R_p=0.71\pm0.02~\mathrm{R_{J}}$) with a wide separation white dwarf companion. TOI-5293 A ($J=12.47$) is an M3 dwarf hosting a short-period ($2.930289 \pm 0.000004$ days) gas giant ($M_p=0.54\pm0.07~\mathrm{M_{J}}$ and $R_p=1.06\pm0.04~\mathrm{R_{J}}$) with a wide separation M dwarf companion. We characterize both systems using a combination of ground-based and space-based photometry, speckle imaging, and high-precision radial velocities from the Habitable-zone Planet Finder and NEID spectrographs. TOI-3984 A b ($T_{eq}=563\pm15$ K and $\mathrm{TSM}=138_{-27}^{+29}$) and TOI-5293 A b ($T_{eq}=675_{-30}^{+42}$ K and $\mathrm{TSM}=92\pm14$) are two of the coolest gas giants among the population of hot Jupiter-sized gas planets orbiting M dwarfs and are favorable targets for atmospheric characterization of temperate gas giants and three-dimensional obliquity measurements to probe system architecture and migration scenarios.

\end{abstract}



\section{Introduction} 
Hot Jupiters (defined as $P<10$ days and $R_{p}\gtrsim8~\mathrm{R_{\oplus}}$ in this work) are a rare class of exoplanets with an occurrence rate of \(\lesssim 1\%\) \cite[see][and references therein]{Beleznay2022} around Sun-like stars measured through radial velocity (RV) surveys \citep[e.g.,][]{Cumming2008,Mayor2011,Wright2012} and photometric surveys \citep[e.g.,][]{Howard2012,Petigura2018,Zhou2019a}.  While the exact formation process is unknown \citep[see the review by][]{Dawson2018}, the most promising channels for origins are \textit{in-situ} formation \citep[e.g.,][]{Boley2016,Batygin2016} or \textit{ex-situ} formation with gas disk migration \citep[e.g.,][]{Lin1996} or high eccentricity tidal migration \citep[e.g.,][]{Rasio1996,Weidenschilling1996,Ford2008,Petrovich2015}. 

Less is known about the formation of hot Jupiters orbiting M dwarfs, which should be difficult to form under the process of core-accretion \citep[e.g.,][]{Laughlin2004,Ida2005,Kennedy2008}. Surveys of M dwarfs have revealed that: (i) small ($R_p<4\mathrm{~R_\oplus}$) planets on short-period orbits ($P<200$ days) are more common around M dwarfs than Sun-like stars \citep[e.g.,][]{Dressing2015,Mulders2015,Hardegree-Ullman2019,Hsu2020} and (ii) low-mass planets ($1\mathrm{~M_\oplus}<M_p<10\mathrm{~M_\oplus}$) are more frequently found orbiting later type M dwarfs on short-period orbits ($P<200$ days) \citep[e.g.,][]{Bonfils2013,Tuomi2014,Tuomi2019,Sabotta2021,Pinamonti2022}. The majority of the aforementioned surveys have not detected any appreciable number of hot Jupiters and can only place upper limits of $\sim2\%$ on the intrinsic occurrence rate. 

A tighter constraint on the occurrence of hot Jupiters orbiting M dwarfs has been made possible through an analysis of M dwarfs observed in the primary TESS mission \citep{Ricker2015}. \cite{Gan2023} report an occurrence rate of $0.27\pm0.09\%$ for hot Jupiters orbiting early M dwarfs from an analysis of 60819 M dwarfs with $10.5<T<13.5$ spanning $0.45~M_\odot<M_\star<0.65~M_\odot$. \cite{Bryant2023} independently derived an occurrence rate of $0.194\pm0.072\%$ from an analysis of 91306 M dwarfs spanning $0.088~M_\odot<M_\star<0.71~M_\odot$. At present, these values are smaller than occurrence rates for hot Jupiters orbiting Sun-like stars, although they are consistent within $1-3\sigma$. An analysis of a larger sample of TESS M dwarfs is needed to refine the occurrence rate and determine whether the M dwarf gas giant population is consistent with the population orbiting Sun-like stars.

In this paper, we confirm the planetary nature of two gas giants transiting the M dwarfs TOI-3984 A ($J=11.93$, $T=13.46$) and TOI-5293 A ($J=12.47$, $T=13.98$). We characterize these system using space and ground-based photometry, speckle imaging, and precision RVs obtained with the Habitable-zone Planet Finder \citep[HPF;][]{Mahadevan2012,Mahadevan2014} and NEID \citep{Schwab2016,Halverson2016} spectrographs. We derive stellar parameters for the host stars with HPF spectra and jointly model the photometry and RVs to confirm the planetary nature of TOI-3984 A b and TOI-5293 A b.

This paper is structured as follows: Section \ref{sec:observations} presents the photometric, imaging, and spectroscopic observations. The best estimates of the stellar parameters and properties are presented in Section \ref{sec:stellarpar}. The modeling and analysis of the photometry and RVs are presented in Section \ref{sec:modelfit}. Section \ref{sec:discussion} provides further discussion of the nature of these planets and the feasibility for future study. A summary of our key results is presented in Section \ref{sec:summary}.

\section{Observations} \label{sec:observations}
\subsection{TESS}
TESS \citep{Ricker2015} observed (i) TOI-3984 A (TIC 20182780, Gaia DR3 1291955578869575552) in long-cadence mode (30-min cadence) during Sectors $23-24$ (2020 March 18 through 2020 May 13) and in short-cadence mode (2-min cadence) during Sectors $50-51$ (2022 April 22 through 2022 May 18) and (ii) TOI-5293 A (TIC 250111245, Gaia DR3 2640121486388076032) in long-cadence mode during Sector 42 (2021 August 20 through 2021 September 16). Each star has one planet candidate, TOI-3984.01 and TOI-5293.01, that was identified by the ``quick-look pipeline'' \citep[QLP\footnote{\url{https://tess.mit.edu/qlp/}};][]{Huang2020,Huang2020b} as part of a search for planet candidates orbiting stars with a TESS magnitude of $T>12$ \citep{Kunimoto2022}. 

\begin{figure*}[!ht]
\epsscale{1.2}
\plotone{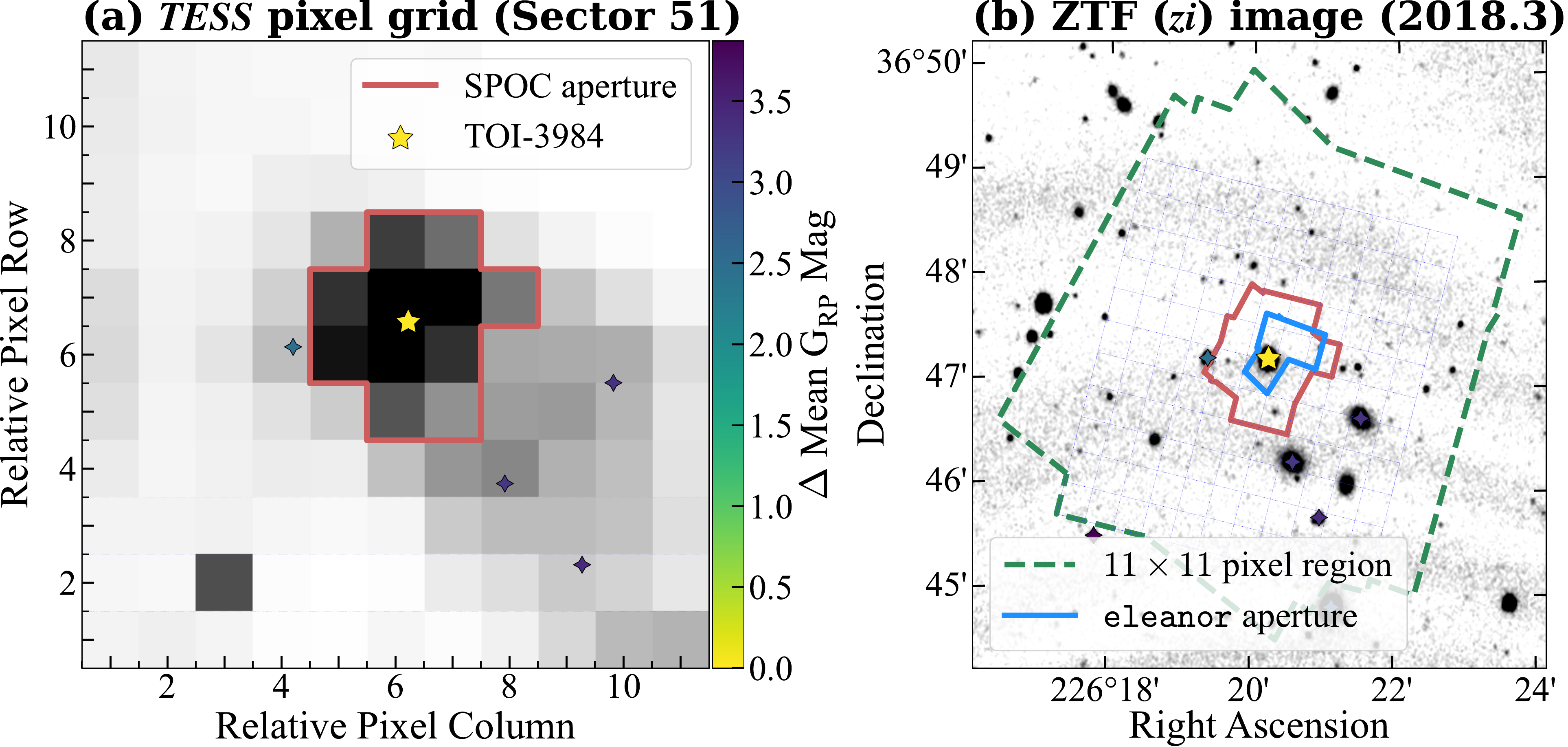}
\caption{\textbf{(a)} The $11\times11$ TESS target pixel cutout from Sector 51 centered on TOI-3984 A (marked as a star). Stars identified in Gaia DR3 with magnitudes $\Delta\mathrm{G_{RP}}<4$ are marked with diamond stars for reference. \textbf{(b)} Overlay of the region contained in an $11\times11$ pixel grid centered on TOI-3984 A from all sectors (green polygon), the region contained within all SPOC (red polygon) and \texttt{eleanor} (blue polygon) photometric apertures, and other comparably bright stars on a ZTF $zi$ image.}\vspace{0.21cm}
\label{fig:3984apertures}
\end{figure*}

\begin{figure*}[!ht]
\epsscale{1.2}
\plotone{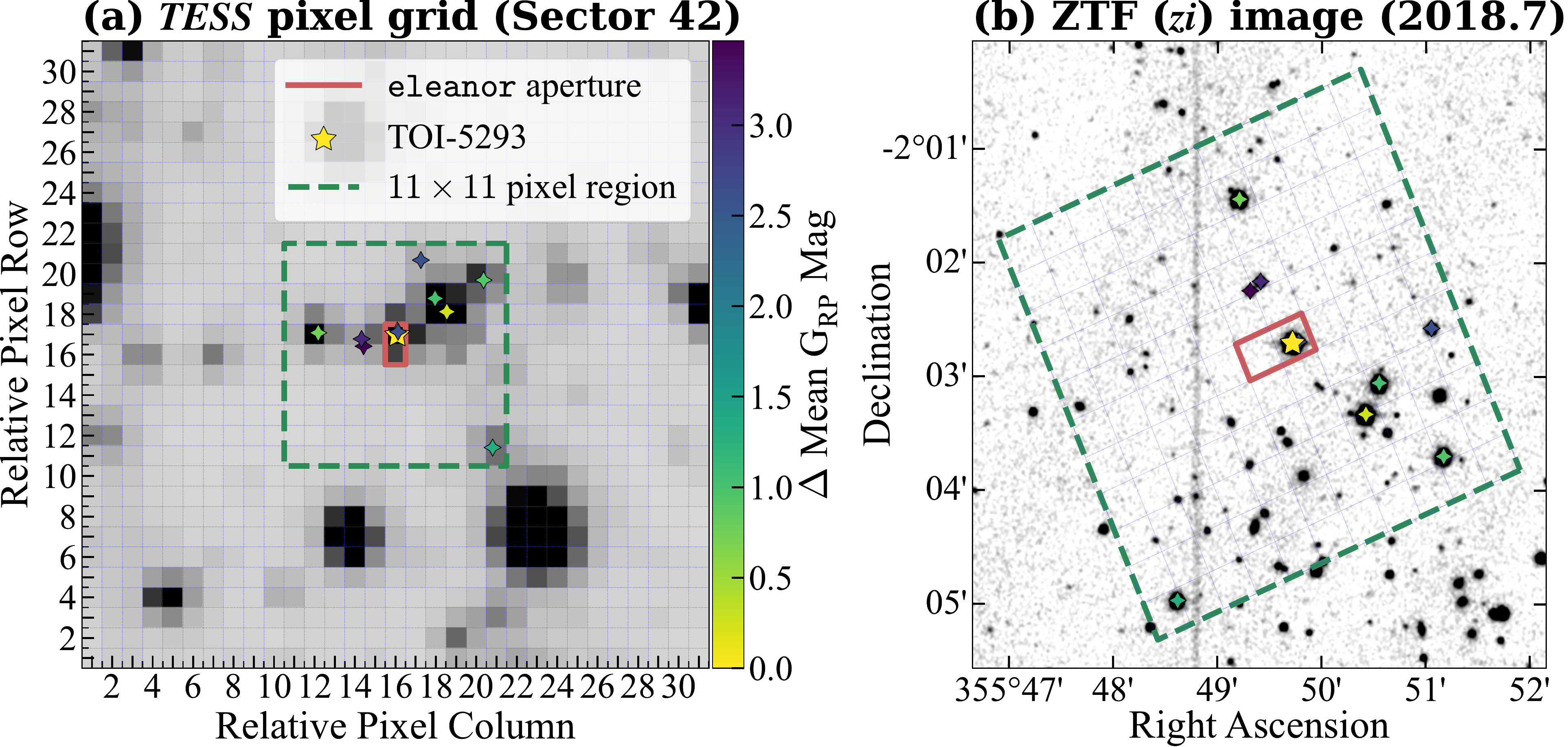}
\caption{Identical to Figure \ref{fig:3984apertures} but for TOI-5293 A. \textbf{(a)} The $31\times31$ TESS target pixel cutout centered around TOI-5293 A (marked as a star). Stars identified in Gaia EDR3 having magnitudes $\Delta\mathrm{G_{RP}}<4$ are marked with diamond stars. \textbf{(b)} Overlay of the TESS $11\times11$ pixel subgrid (green polygon), the region contained in the photometric aperture (red polygon), and other comparably bright stars on a ZTF $zi$ image.}\vspace{0.21cm}
\label{fig:5293apertures}
\end{figure*}

We extract the long-cadence photometry from the TESS full-frame images using \texttt{eleanor} \citep{Feinstein2019} to process a cut-out of \(31\times31\) pixels from the calibrated full-frame images centered on each target. The light curves from \texttt{eleanor}\footnote{\href{https://github.com/afeinstein20/eleanor}{https://github.com/afeinstein20/eleanor~\faGithub}} use an aperture which minimizes the hour-binned combined differential photometric precision \citep[CDPP; see][]{Jenkins2010} to ensure that sharp features on timescales of a few hours, such as transits, are preserved. Figures \ref{fig:3984apertures} and \ref{fig:5293apertures} present the apertures used to extract the TESS light curves for TOI-3984 A and TOI-5293 A, respectively. In each figure, panel (a) presents the apertures used to derive the light curve from the latest TESS sector. In panel (b), the region contained in an 11x11 pixel subgrid and the photometric apertures from all TESS sectors are overlaid on images from the Zwicky Transient Facility \citep[ZTF;][]{Bellm2019,Graham2019,Masci2019}.

TOI-5293 A has a nearby companion in the optimal aperture (TIC 2052711961, $T=18.44$, $\Delta~\mathrm{G_{RP}}=2.6$) at a separation of $3.57\arcsec$. The second release of the Pan-STARRS survey \citep[PS1;][]{Chambers2016,Magnier2020} measures a PS1 $i^\prime=17.15$ for the nearby companion, which is too faint to contribute to any significant dilution in the photometry obtained with TESS or small ground-based telescopes described in Sections \ref{obs:rbo}, \ref{obs:tmmt}, and \ref{obs:lcobot}. 

The long-cadence light curves analyzed in this work are the \texttt{CORR\_FLUX} values produced by \texttt{eleanor}. The combined differential photometric precision is the value minimized by \texttt{eleanor} when searching for the best aperture and is 5661 ppm for TOI-3984 A and 7089 ppm for TOI-5293 A.  Observations where the background flux exceeds a threshold (\texttt{FLUX\_BKG} $>2.5\times$\texttt{CORR\_FLUX}) or with non-zero data quality flags \citep[see a detailed description in Table 28 in][]{Tenenbaum2018} are excluded from further analysis. 

\begin{figure*}[!ht]
\epsscale{1.15}
\plotone{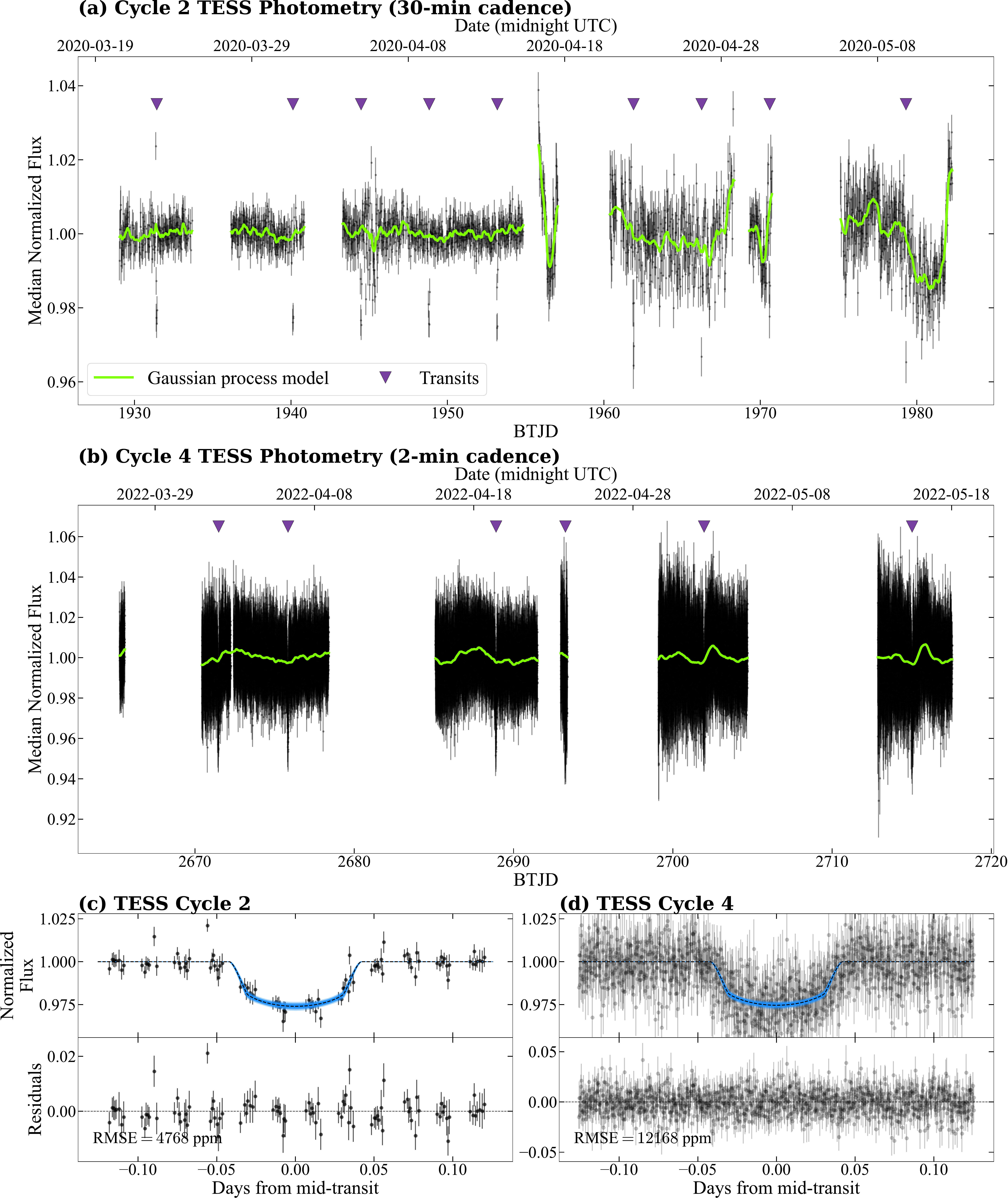}
\caption{\textbf{(a)} The median normalized TESS long-cadence light curve for TOI-3984 A derived with \texttt{eleanor}. The solid green line is the best-fitting Gaussian process model used to detrend the light curve. The triangles indicate the observed transits. \textbf{(b)} Identical to (a) but for short-cadence TESS data. \textbf{(c)} and \textbf{(d)} are the phase-folded light curves for short and long cadence TESS data. In (c) and (d), the best-fitting model from the joint fit to the photometry and RVs is plotted as a dashed line while the shaded regions denote the \(1\sigma\) (darkest), \(2\sigma\), and \(3\sigma\) (lightest) extent of the model posteriors. The modeling of the photometry and RVs is described in detail in Section \ref{sec:modelfit}.}
\label{fig:3984tess}
\end{figure*}

\begin{figure*}[!ht]
\epsscale{1.15}
\plotone{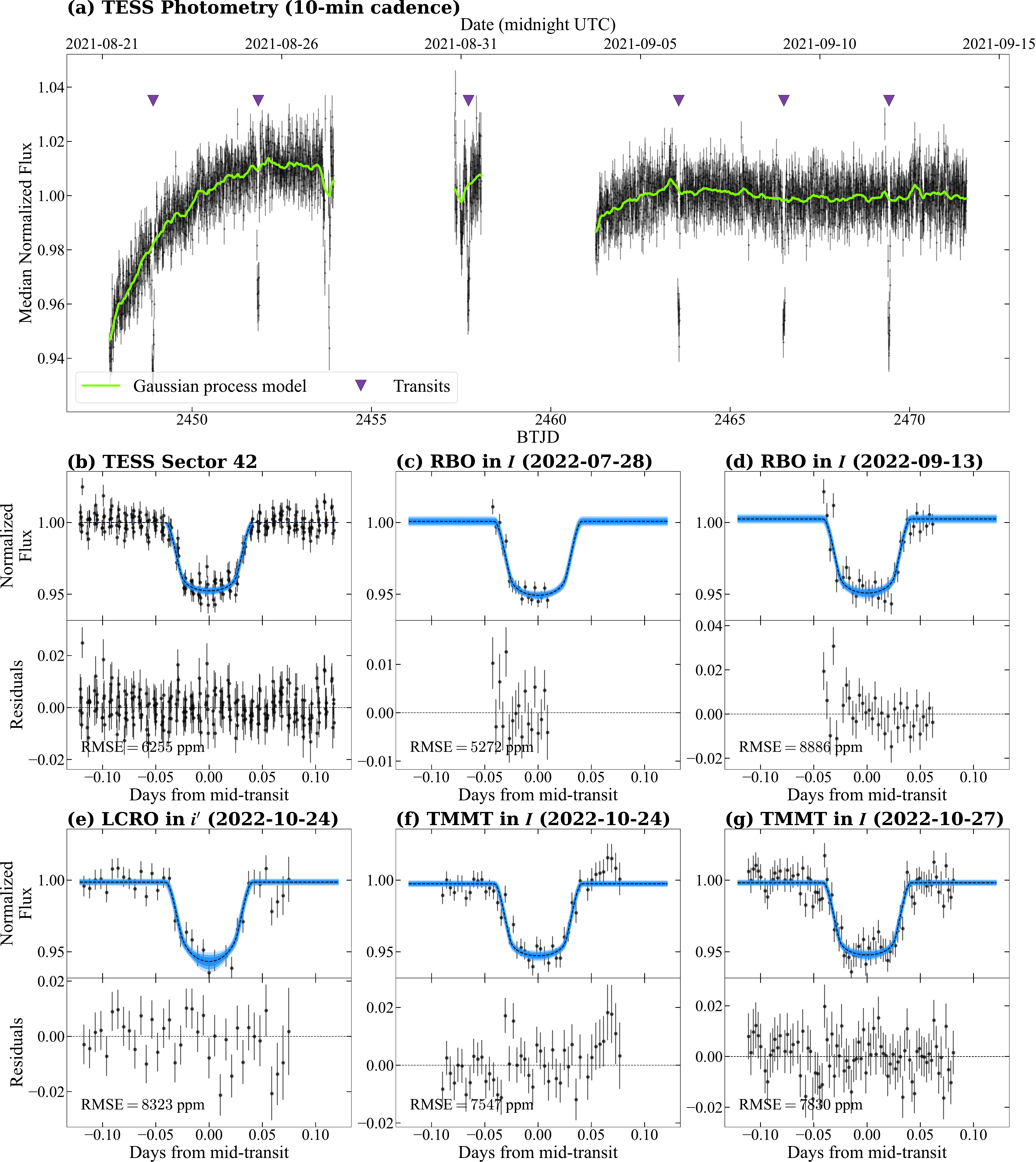}
\caption{Identical to Figure \ref{fig:3984tess}, but for TOI-5293 A. \textbf{(a)} The median normalized TESS light curve for TOI-5293 A derived with \texttt{eleanor} along with the best-fitting Gaussian process model. The triangles indicate the observed transits. \textbf{(b)}$-$\textbf{(g)} are the light curves for the TESS, RBO, LCRO, and TMMT plotted with model posteriors (shaded regions) from the joint fit to the photometry and RVs.}\vspace{1.9cm}
\label{fig:5293tess}
\end{figure*}

The short-cadence photometry of TOI-3984 A is obtained from the pre-search data-conditioned simple aperture photometry \citep[PDCSAP;][]{Jenkins2016} light curves available at the Mikulski Archive for Space Telescopes (MAST)\footnote{\url{https://archive.stsci.edu/prepds/tess-data-alerts/}}. The PDCSAP photometry is corrected for instrumental systematics and dilution from other objects contained within the aperture using algorithms developed for the \textit{Kepler} mission \citep[see][]{Stumpe2012,Smith2012}. As with the long-cadence photometry, observations with non-zero data quality flags are excluded from further analysis. We perform no additional outlier rejection beyond the data quality flags and application of a background threshold value for long-cadence data. Figures \ref{fig:3984tess} and \ref{fig:5293tess} display the TESS light curves. 

\subsection{Ground-based photometry}
\begin{figure*}[!ht]
\epsscale{1.15}
\plotone{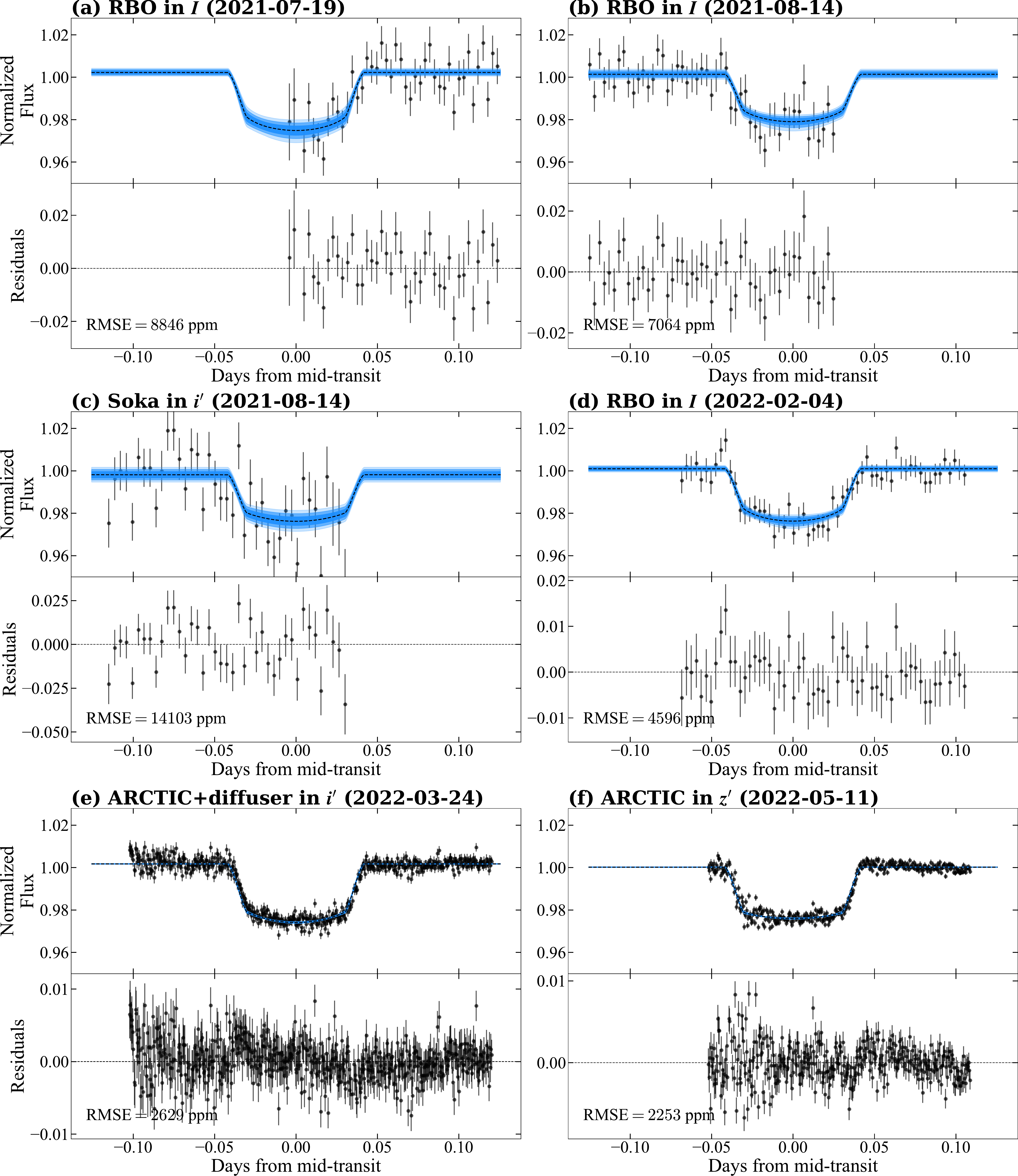}
\caption{Identical to Figure \ref{fig:3984tess}, but displaying the ground-based photometry for TOI-3984 A. \textbf{(a)}$-$\textbf{(f)} are the light curves from RBO, Soka, and ARCTIC plotted with model posteriors (shaded regions) from the joint fit to the photometry and RVs.}\vspace{1.9cm}
\label{fig:3984groundphot}
\end{figure*}

We observed 6 transits of TOI-3984 A and 5 transits of TOI-5293 A using a combination of five separate ground-based facilities. All ground-based light curves were derived using \texttt{AstroImageJ} \citep{Collins2017}. Following the methodology in \cite{Stefansson2017}, the estimated scintillation noise was included in the flux uncertainty. The light curves are not detrended with any external parameter (e.g., airmass or time). The observations are described in detail below and summarized in Table \ref{tab:photometry}. The ground-based photometry for TOI-5293 A and TOI-3984 A are presented in Figures \ref{fig:5293tess} and \ref{fig:3984groundphot}, respectively.

\begin{deluxetable*}{ccccc}
\tablecaption{Summary of ground-based photometry \label{tab:photometry}}
\centering
\tablehead{\colhead{Civil Night} &
\colhead{Facility} & 
\colhead{Filter} &
\colhead{Exposure Time} &
\colhead{Aperture and sky annulus} \\
\colhead{} & 
\colhead{} & 
\colhead{} &
\colhead{(s)} & 
\colhead{(pixels and \arcsec{})} } 
\startdata
\multicolumn{5}{l}{\hspace{-0.2cm} TOI-3984 A:}  \\
~~2021-07-18 & RBO & Bessell I & 240 & 10/20/30 ($7.3/14.6/21.9\arcsec$) \\
~~2021-08-13 & RBO & Bessell I & 240 & 10/20/30 ($7.3/14.6/21.9\arcsec$) \\
~~2021-08-13 & Soka & Sloan $i^\prime$ & 300 & 10/15/20 ($7.0/10.5/14.0\arcsec$) \\
~~2022-02-03 & RBO & Bessell I & 240 & 10/20/30 ($7.3/14.6/21.9\arcsec$) \\
~~2022-03-23 & ARCTIC & Sloan $i^\prime$ & 45 & 20/30/40 ($9.1/13.7/18.2\arcsec$) \\
~~2022-05-10 & ARCTIC & Sloan $z^\prime$ & 30 & 12/20/30 ($5.5/9.1/13.7\arcsec$) \\
\hline
\multicolumn{5}{l}{\hspace{-0.2cm} TOI-5293 A:}  \\
~~2022-07-27 & RBO & Bessell I & 240 & 10/20/30 ($7.3/14.6/21.9\arcsec$) \\
~~2022-09-12 & RBO & Bessell I & 240 & 10/20/30 ($7.3/14.6/21.9\arcsec$) \\
~~2022-10-23 & LCRO & Sloan $i^\prime$ & 420 & 6/10/15 ($4.6/7.7/11.6\arcsec$) \\
~~2022-02-03 & TMMT & Bessell I & 300 & 6/10/15 ($7.2/11.9/17.9\arcsec$) \\
~~2022-03-23 & TMMT & Bessell I & 180 & 6/10/15 ($7.2/11.9/17.9\arcsec$) \\
\enddata
\end{deluxetable*}

\subsubsection{RBO 0.6 m telescope}\label{obs:rbo}
The 0.6 m telescope at the Red Buttes Observatory (RBO) in Wyoming \citep{Kasper2016} is a f/8.43 Ritchey-Cretien constructed by DFM Engineering, Inc. and equipped with an Apogee Alta F16M camera. We observed (i) TOI-3984 A on the nights of 2021 July 18, 2021 August 13, and 2022 February 03 and (ii) TOI-5293 A on the nights of 2022 July 27 and 2022 September 12. All observations were moderately defocused, used an exposure time of 240s, and operated with the Bessell I filter \citep{Bessell1990} and the $2 \times 2$ on-chip binning mode, which provides a gain of 1.39 $\mathrm{e^-/ADU}$, a plate scale of \(0.731 \arcsec/\mathrm{pixel}\), and a readout time of $\sim2.4$s. 

\subsubsection{SUA Nieves 0.3 m telescope}
The 0.3 m Planewave CDK14 telescope located at the Luis and Linda Nieves Observatory\footnote{\url{https://sites.soka.edu/SUO/about/}} at Soka University of America in Aliso Viejo, CA uses a FLI Proline 16803 camera with a $47\times47\arcmin$ field-of-view and a quantum efficiency $>50\%$. We observed TOI-3984 A on the night of 2021 August 13 in the Sloan $i^\prime$ filter using a 300s exposure time and $1\times1$ binning mode, which provides a plate scale of $0.7\arcsec/\mathrm{pixel}$. 

\subsubsection{APO 3.5m telescope}
We used the 3.5m Astrophysical Research Consortium (ARC) Telescope Imaging Camera \citep[ARCTIC;][]{Huehnerhoff2016} on the ARC 3.5m Telescope at Apache Point Observatory (APO) to obtain transits of TOI-3984 A on the nights of 2022 March 23 and 2022 May 10. The observation on 2022 March 23 were performed in the Sloan $i^\prime$ filter using an engineered diffuser \citep{Stefansson2017} with an exposure time of 45 s. ARCTIC was operated in the quad-amplifier and fast readout modes using the $4 \times 4$ on-chip binning mode to achieve a gain of 2 $\mathrm{e^-/ADU}$, a plate scale of $0.456 \mathrm{\arcsec/pixel}$, and a readout time of 1.3 s. The observation on 2022 May 10 was performed in the Sloan $z^\prime$ filter slightly out of focus with an exposure time of 30s. Hardware issues prevented the use of quad mode and instead ARCTIC was operated in the single-amplifier and fast readout modes using the $4 \times 4$ on-chip binning mode with a readout time of 11 s. 

\subsubsection{LCO 0.3m TMMT telescope}\label{obs:tmmt}
The robotic Three-hundred MilliMeter Telescope \citep[TMMT;][]{Monson2017} at Las Campanas Observatory (LCO) is a f/7.8 FRC300 from Takahashi on a German equatorial AP1600 GTO mount with an Apogee Alta U42-D09 CCD Camera, FLI ATLAS focuser, and Centerline filter wheel. We observed TOI-5293 A on the nights of 2022 October 23 and 2022 October 26. The observations were performed slightly out of focus in the Bessell $I$ filter \citep{Bessell1990} with an exposure time of 300s and 180s, respectively, while operating in a 1x1 binning mode. In this mode, TMMT has a gain of 1.35 $\mathrm{e^-/ADU}$, a plate scale of \(1.194 \arcsec/\mathrm{pixel}\), and a readout time of $\sim6$s. 

\subsubsection{LCRO 0.3m telescope}\label{obs:lcobot}
The 305 mm Las Campanas Remote Observatory telescope\footnote{\url{http://lcobot.duckdns.org/}} (LCRO) at LCO observed TOI-5293 A on the night of 2022 October 23. LCRO is an f/8 Maksutov-Cassegrain from Astro-Physics on a German Equatorial AP1600 GTO mount with an FLI Proline 16803 CCD Camera, FLI ATLAS focuser and Centerline filter wheel. The observations were performed slightly out of focus in the Sloan $i'$ filter with an exposure time of 420s. We used the $1 \times 1$ binning mode, providing a gain of 1.52 $\mathrm{e^-/ADU}$, a plate scale of \(0.773 \arcsec/\mathrm{pixel}\), and a readout time of 17s. 

\subsection{High-contrast imaging}
NESSI \citep{Scott2018} is a dual-channel speckle imager on the WIYN 3.5m Telescope at Kitt Peak National Observatory (KPNO). Both TOI-3984 A and TOI-5293 A were observed on 2022 April 18 using NESSI. The faintness of these targets ($r^\prime>14$) prevented observations in the narrow filters that NESSI traditionally uses, while hardware issues during the observing run only allowed for observing with the redder channel in the Sloan \(z^\prime\) filter. The images were reconstructed following the procedures described in \cite{Howell2011}.

\begin{figure*}[!ht]
\epsscale{1.15}
\plotone{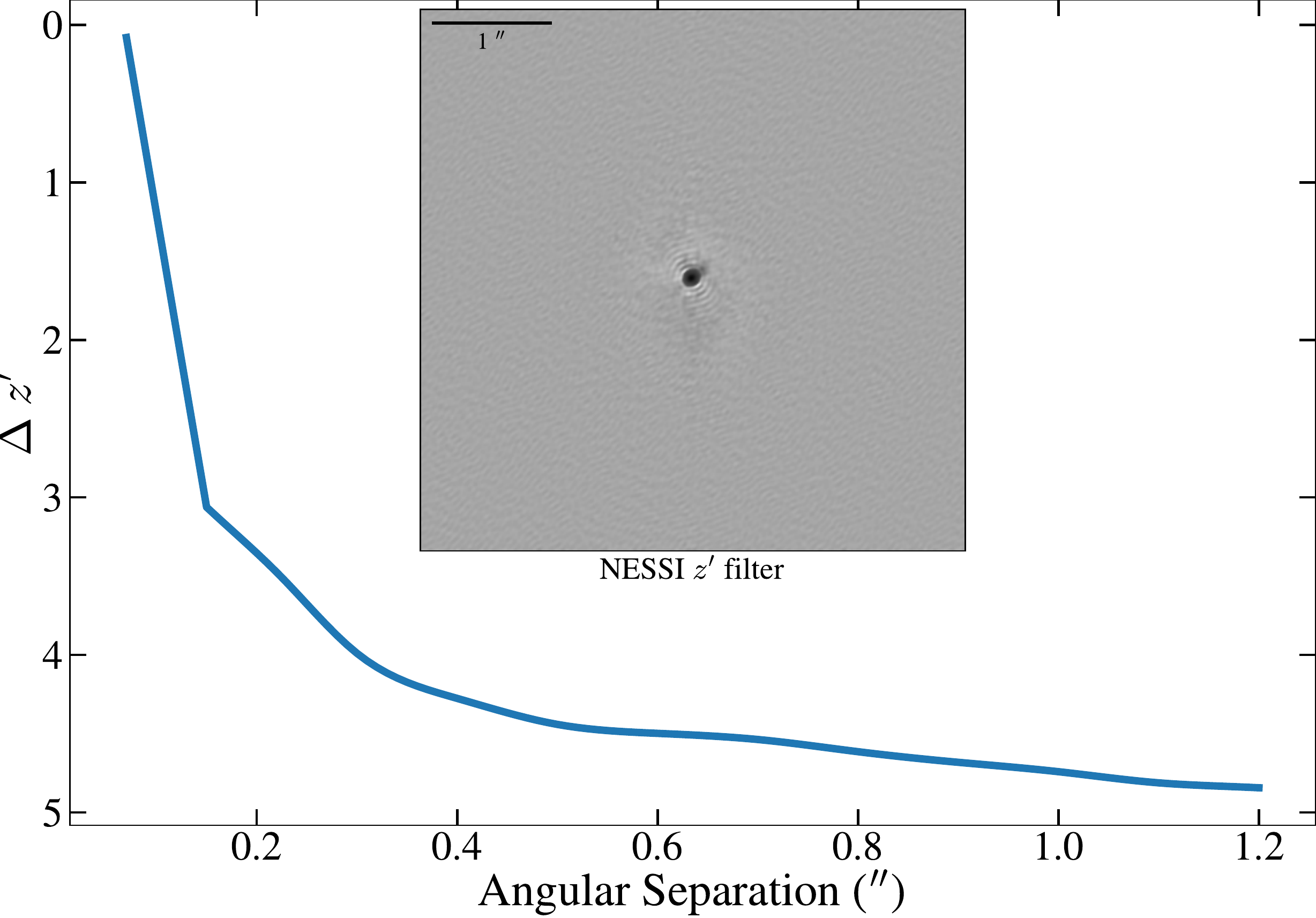}
\caption{The \(5\sigma\) contrast curves for TOI-3984 A obtained from speckle imaging with NESSI in the Sloan \(z^\prime\) filter. The data reveal no bright companions at separations of $0.2\arcsec{}-1.75\arcsec{}$. The inset is the $4.7\arcsec{}\times4.7\arcsec{}$ NESSI speckle image centered on TOI-3984 A in the Sloan \(z^\prime\) filter.}
\label{fig:3984imaging}
\end{figure*}

\begin{figure*}[!ht]
\epsscale{1.15}
\plotone{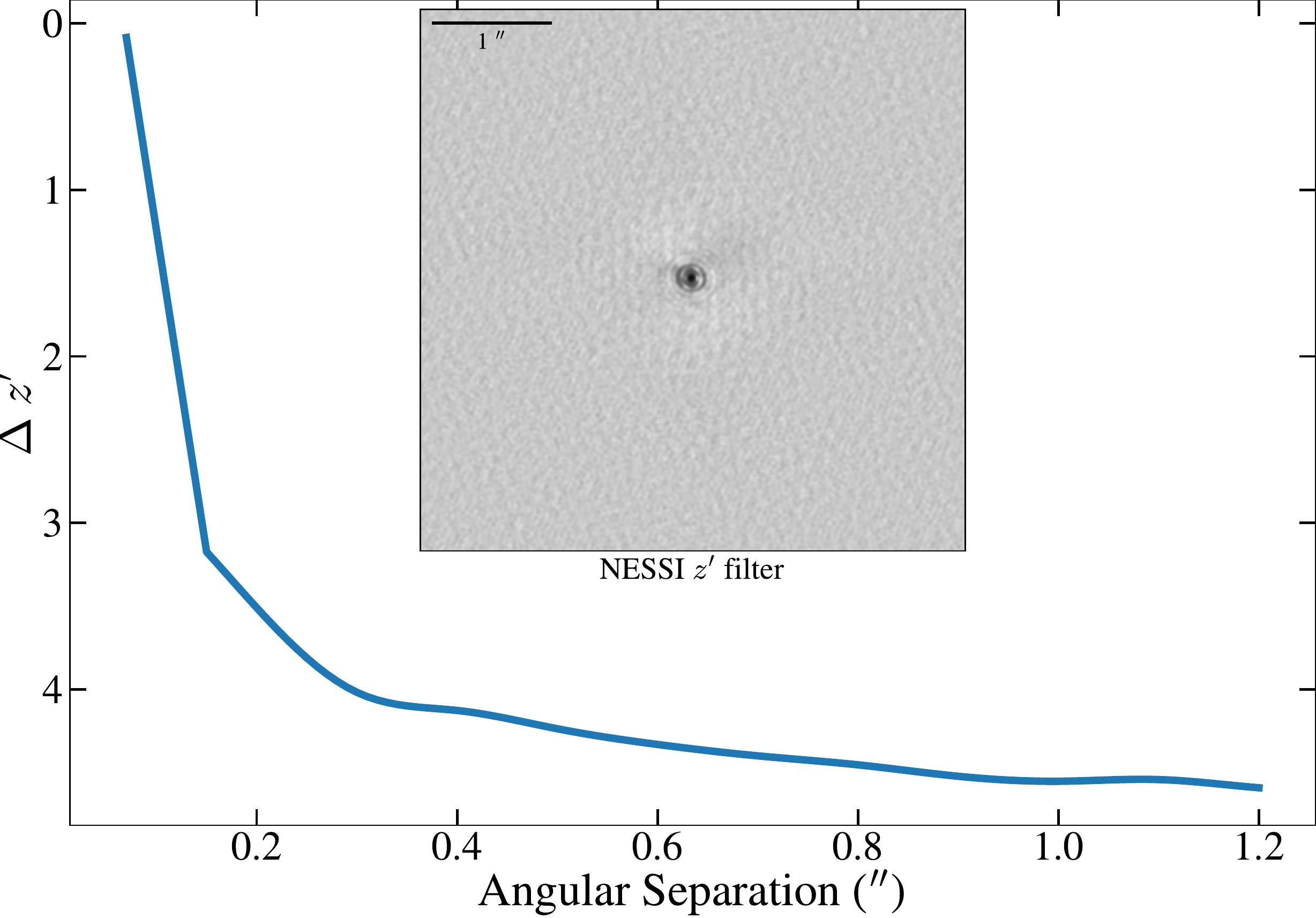}
\caption{Similar to Figure \ref{fig:3984imaging} but for TOI-5293 A.  The \(5\sigma\) contrast curve for TOI-5293 A obtained from speckle imaging using NESSI along with an inset displaying the $4.7\arcsec{}\times4.7\arcsec{}$ reconstructed speckle image in the Sloan \(z^\prime\) filter.}
\label{fig:5293imaging}
\end{figure*}

Figures \ref{fig:3984imaging} and \ref{fig:5293imaging} display the $5\sigma$ contrast curve with insets of the reconstructed NESSI speckle images. The NESSI data reveal no bright ($\Delta z^\prime < 3$) companions and no potential sources of dilution at separations of $0.2-1.2\arcsec$ from the host stars. These angular limits correspond to projected spatial limits of $22-130$ au for TOI-3984 A and $32-193$ au for TOI-5293 A. AO imaging of TOI-3984 A \citep[see][]{Gan2023} obtained with the Palomar High Angular Resolution Observer on the Palomar-5.1m telescope \citep{Hayward2001} also reveal it is an isolated star having a contrast of $\Delta\textrm{mag}>6.1$ at distances $>0.5\arcsec{}$ ($>54$ au) from the star in the near-infrared ($1-2.5$ \textmu{}m).

\subsection{High-precision spectroscopy}

\subsubsection{HPF spectrograph}
HPF \citep{Mahadevan2012,Mahadevan2014} is a high-resolution ($R\sim55,000$), fiber-fed \citep{Kanodia2018a}, temperature stabilized \citep{Stefansson2016},  near-infrared (\(\lambda\sim8080-12780\)\ \AA) spectrograph located on the 10m Hobby-Eberly Telescope at McDonald Observatory \citep[HET;][]{Ramsey1998,Hill2021}. Observations are executed in a queue by the HET resident astronomers \citep{Shetrone2007}. We obtained (i) 35 visits of TOI-3984 A between 2021 August 23 and 2022 May 10 and (ii) 16 visits of TOI-5293 A between 2022 September 10 and 2022 December 01 with median signal-to-noise ratios (S/N) per 1D extracted pixel at 1070nm of 38 and 35, respectively.

The \texttt{HxRGproc} tool\footnote{\href{https://github.com/indiajoe/HxRGproc}{https://github.com/indiajoe/HxRGproc~\faGithub}} \citep{Ninan2018} processed the raw HPF data and performed bias noise removal, nonlinearity correction, cosmic-ray correction, and slope/flux and variance image calculations. The 1D spectra were extracted following the procedures in \cite{Ninan2018}, \cite{Kaplan2019}, and \cite{Metcalf2019}. The wavelength solution and drift correction were extrapolated using laser frequency comb (LFC) frames obtained from routine calibrations, which enable wavelength calibration on the order of $<30~\mathrm{cm~s^{-1}}$ \citep[see Appendix A in][]{Stefansson2020}. 

The RVs were calculated using a modified version of the \texttt{SpEctrum Radial Velocity AnaLyser} code \citep[\texttt{SERVAL};][]{Zechmeister2018} optimized for HPF RV extractions (see \cite{Metcalf2019} and \cite{Stefansson2020} for details). \texttt{SERVAL} uses the template-matching technique to derive RVs \citep[e.g.,][]{Anglada-Escude2012} and creates a master template from the observations by minimizing the \(\chi^2\) statistic. The master template is generated from the observed spectra after masking sky-emission lines and telluric regions identified using a synthetic telluric-line mask generated from \texttt{telfit} \citep{Gullikson2014}. The barycentric correction is calculated using \texttt{barycorrpy} \citep{Kanodia2018}. Table \ref{tab:rvs} reports the HPF RVs, the \(1\sigma\) uncertainties, and the S/N per pixel for TOI-3984 A and TOI-5293 A. Figure \ref{fig:3984rv} presents the RVs for TOI-3984 A and Figure \ref{fig:5293rv} presents the RVs for TOI-5293 A.

\subsubsection{NEID spectrograph}
\begin{figure*}[!ht]
\epsscale{1.15}
\plotone{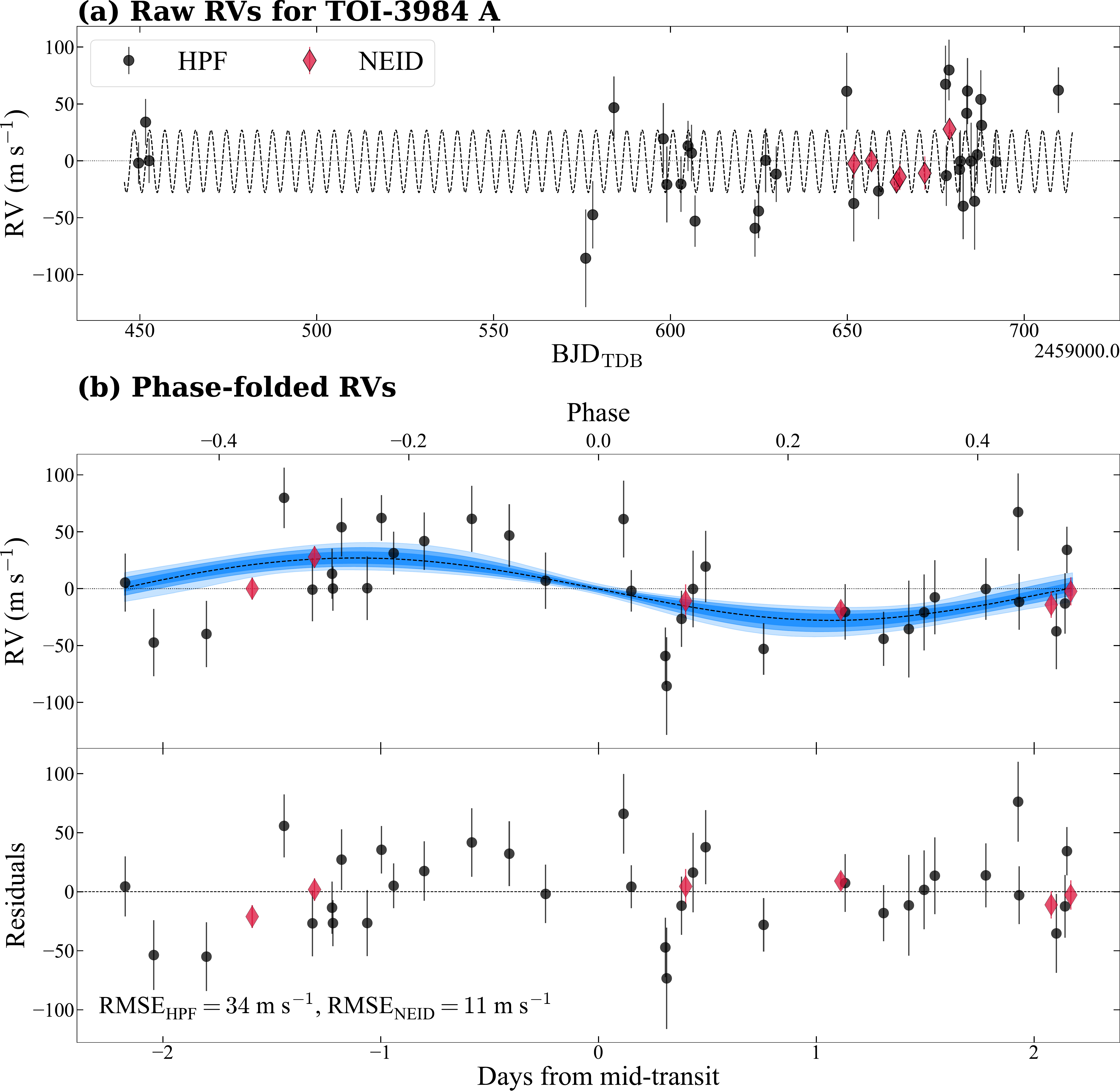}
\caption{\textbf{(a)} shows the RVs for TOI-3984 A derived with modified versions of \texttt{SERVAL}. \textbf{(b)} displays the phase-folded RVs plotted with model posteriors and the residuals to the fit. In (a) and (b), the dashed line is the best-fitting Keplerian model. The shaded regions denote the \(1\sigma\) (darkest), \(2\sigma\), and \(3\sigma\) (lightest) extent of the model posteriors. The modeling is described in Section \ref{sec:modelfit}.}
\label{fig:3984rv}
\end{figure*}

\begin{figure*}[!ht]
\epsscale{1.15}
\plotone{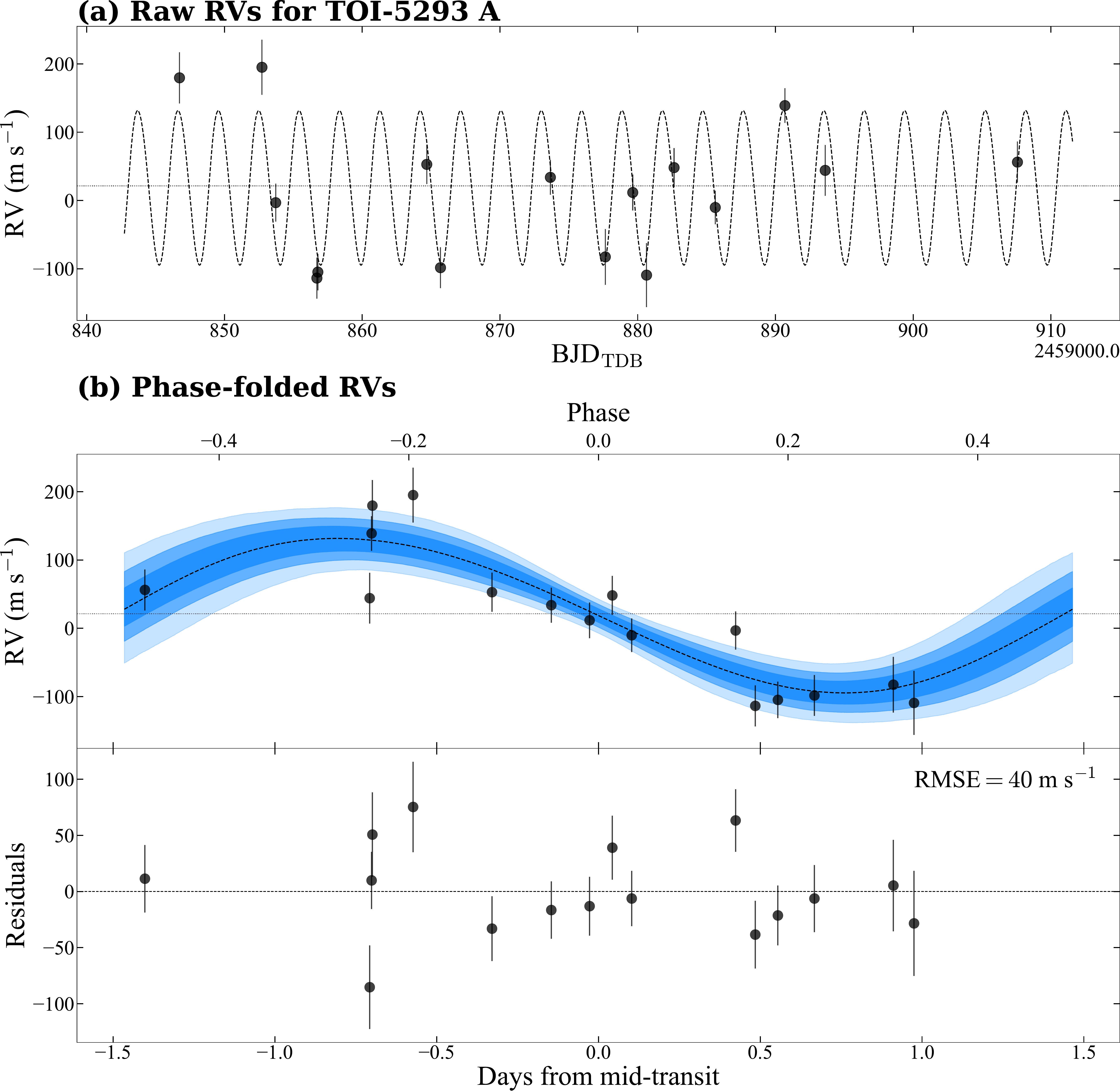}
\caption{Identical to Figure \ref{fig:3984rv}, but for TOI-5293 A. \textbf{(a)} shows the HPF RVs. \textbf{(b)} displays the phase-folded RVs plotted with model posteriors.}
\label{fig:5293rv}
\end{figure*}

NEID \citep[][]{Schwab2016,Halverson2016} is an environmentally stabilized \citep{Stefansson2016,Robertson2019}, high-resolution ($R\sim110,000$), fiber-fed \citep{Kanodia2018a} spectrograph with an extended red wavelength coverage (\(\lambda\sim3800-9300\)\ \AA) installed on the WIYN 3.5m telescope at KPNO in Arizona. We obtained 6 visits of TOI-3984 A between 2022 March 13 and 2022 April 09 in high-resolution mode with a median S/N per 1D extracted pixel of 12 at 850nm. The NEID data were reduced using the NEID Data Reduction Pipeline\footnote{\url{https://neid.ipac.caltech.edu/docs/NEID-DRP/}} (DRP), and the Level-2 1D extracted spectra were retrieved from the NEID Archive\footnote{\url{https://neid.ipac.caltech.edu/}}. We measured the RVs using a modified version of the \texttt{SERVAL} code \cite[see][]{Stefansson2022} and extracted RVs using the wavelength range from $5440-8920$ \AA{} (order indices $61-104$) and the inner most 3000 pixels of each order \citep[similar to][]{Canas2022b}. Figure \ref{fig:3984rv} presents the RVs while Table \ref{tab:rvs} reports the NEID RVs, the \(1\sigma\) uncertainties, and the S/N per pixel.

\startlongtable
\begin{deluxetable}{lrcccc}
\tabletypesize{\scriptsize }
\tablecaption{Radial velocities. \label{tab:rvs}}
\tablehead{
\colhead{$\mathrm{BJD_{TDB}}$}  &
\colhead{RV} &
\colhead{$\sigma$} & 
\colhead{S/N$^a$} &
\colhead{Instrument}\\
& 
\colhead{$(\mathrm{m~s^{-1}})$} & 
\colhead{$(\mathrm{m~s^{-1}})$} & 
 &
}
\startdata
\multicolumn{5}{l}{\hspace{-0.2cm} TOI-3984 A:}  \\
~~2459449.620693 &    0.5 &  18.2 & 49 & HPF \\
~~2459451.620722 &   36.5 &  20.4 & 45 & HPF \\
~~2459452.603981 &    2.5 &  19.5 & 47 & HPF \\
~~2459574.028699 &  $ -78.2 $ &  89.8 & 34 & HPF \\
~~2459576.029258 &  $ -83.2 $ &  43.0 & 34 & HPF \\
~~2459578.027929 &  $ -45.0 $ &  29.6 & 34 & HPF \\
~~2459584.013323 &   49.1 &  27.5 & 36 & HPF \\
~~2459597.974678 &   21.8 &  31.4 & 31 & HPF \\
~~2459598.978176 &  $ -18.4 $ &  33.4 & 31 & HPF \\
~~2459602.968624 &  $ -18.0 $ &  24.4 & 38 & HPF \\
~~2459604.966266 &   15.5 &  22.1 & 43 & HPF \\
~~2459605.946813 &    9.3 &  24.7 & 38 & HPF \\
~~2459606.947872 &  $ -50.6 $ &  22.7 & 41 & HPF \\
~~2459623.909836 &  $ -56.8 $ &  25.1 & 37 & HPF \\
~~2459624.911946 &  $ -41.7 $ &  23.8 & 40 & HPF \\
~~2459626.894567 &    2.8 &  28.0 & 40 & HPF \\
~~2459629.887569 &   $ -9.2 $ &  24.5 & 40 & HPF \\
~~2459649.837986 &   63.6 &  33.7 & 40 & HPF \\
~~2459651.825176 &  $ -35.1 $ &  33.4 & 40 & HPF \\
~~2459658.810776 &  $ -24.1 $ &  24.7 & 40 & HPF \\
~~2459677.769462 &   69.8 &  33.9 & 36 & HPF \\
~~2459677.984824 &  $ -10.6 $ &  26.6 & 36 & HPF \\
~~2459678.752641 &   82.2 &  26.6 & 36 & HPF \\
~~2459681.740411 &   $ -5.2 $ &  32.6 & 35 & HPF \\
~~2459681.974403 &    2.1 &  27.1 & 35 & HPF \\
~~2459682.749492 &  $ -37.4 $ &  29.1 & 35 & HPF \\
~~2459683.749510 &   44.2 &  25.2 & 38 & HPF \\
~~2459683.967863 &   63.7 &  29.0 & 33 & HPF \\
~~2459684.983417 &    2.2 &  33.7 & 30 & HPF \\
~~2459685.974345 &  $ -33.1 $ &  42.6 & 38 & HPF \\
~~2459686.729534 &    7.7 &  25.5 & 38 & HPF \\
~~2459687.723023 &   56.4 &  25.7 & 38 & HPF \\
~~2459687.962405 &   33.6 &  18.9 & 49 & HPF \\
~~2459691.942885 &    1.5 &  27.9 & 35 & HPF \\
~~2459709.673097 &   64.5 &  20.1 & 47 & HPF \\
~~2459651.891110 &  $-60.1$ &  12.4 & 12 & NEID \\ 
~~2459656.839623 &  $-57.8$ &   9.7 & 14 & NEID \\ 
~~2459663.895380 &  $-76.5$ &   7.1 & 19 & NEID \\ 
~~2459664.861500 &  $-71.9$ &  11.5 & 12 & NEID \\ 
~~2459671.889748 &  $-68.6$ &  14.7 & 10 & NEID \\ 
~~2459678.892545 &  $-29.9$ &   9.6 & 15 & NEID \\ 
\hline
\multicolumn{5}{l}{\hspace{-0.2cm} TOI-5293 A:}  \\
~~2459846.735412 &   179.6 &  37.6 & 28 & HPF \\
~~2459852.721876 &   195.1 &  40.3 & 27 & HPF \\
~~2459853.718534 &    $-3.3$ &  27.9 & 36 & HPF \\
~~2459856.709588 &  $-113.7$ &  30.2 & 35 & HPF \\
~~2459856.779480 &  $-104.9$ &  26.7 & 38 & HPF \\
~~2459864.686793 &    52.8 &  28.8 & 37 & HPF \\
~~2459865.683293 &   $-98.5$ &  29.9 & 35 & HPF \\
~~2459873.660887 &    33.8 &  25.7 & 39 & HPF \\
~~2459877.648474 &   $-82.7$ &  40.8 & 26 & HPF \\
~~2459879.639891 &    11.6 &  26.2 & 38 & HPF \\
~~2459880.642541 &  $-109.4$ &  46.9 & 23 & HPF \\
~~2459882.640435 &    48.1 &  28.5 & 35 & HPF \\
~~2459885.630585 &   $-10.3$ &  24.6 & 41 & HPF \\
~~2459890.687347 &   138.9 &  25.4 & 39 & HPF \\
~~2459893.611705 &    44.1 &  37.3 & 29 & HPF \\
~~2459907.568470 &    56.2 &  30.1 & 34 & HPF \\
\enddata
\tablenotetext{a}{The S/N is the median value per 1D extracted pixel at 1070nm for HPF and 850nm for NEID. HPF observations use an exposure time of 1890s. NEID observations use an exposure time of 1800s.}
\end{deluxetable}

\section{Stellar Parameters}\label{sec:stellarpar}
\subsection{Spectroscopic parameters}\label{sec:specmatch}
The stellar effective temperature ($T_e$), surface gravity ($\log g_\star$), and metallicity ([Fe/H]) were calculated using the \texttt{HPF-SpecMatch}\footnote{\href{https://gummiks.github.io/hpfspecmatch/}{https://gummiks.github.io/hpfspecmatch~\faGithub}} package \citep[][]{Stefansson2020}, which derives stellar parameters using the empirical template matching methodology discussed in \cite{Yee2017}. It uses a two-step \(\chi^{2}\) minimization to identify the five best-matching spectra from a library of well-characterized stars and derives spectroscopic parameters using a weighted, linear combination of the library stars. \texttt{HPF-SpecMatch} derives $v\sin i_\star$ by broadening the library spectra using a linear limb darkening law \citep{Gray2005}. The reported uncertainties are the standard deviation of the residuals from a leave-one-out cross-validation procedure applied to the entire spectral library.

The HPF spectral library contains 166 stars and spans the following parameter space: $2700~\mathrm{K} < T_{e} < 6000~\mathrm{K}$, $4.3<\log g_\star < 5.3$, and $-0.5 < \mathrm{[Fe/H]} < 0.5$. The library includes 87 M dwarfs ($T_{e}\le4000$ K) of which 40 are M dwarfs spanning $3300\mathrm{K} \le T_{e} \le 3700~\mathrm{K}$, $4.66<\log g_\star < 4.97$, and $-0.5 < \mathrm{[Fe/H]} < 0.4$. The spectral matching was performed using HPF order index 5 ($8534-8645$ \AA) due to minimal telluric contamination. TOI-3984 A is determined to have $T_{e}=3476\pm88$ K, $\log g_\star=4.81\pm0.05$, and $\mathrm{[Fe/H]=0.18\pm0.12}$. TOI-5293 A is determined to have $T_{e}=3586\pm88$ K, $\log g_\star=4.77\pm0.05$, and $\mathrm{[Fe/H]=-0.03\pm0.12}$. The resolution limit of HPF ($R\sim55,000$) places a constraint of $v \sin i < 2 \mathrm{~km~s^{-1}}$ for both TOI-3984 A and TOI-5293 A. Table \ref{tab:stellarparam} presents the derived spectroscopic parameters with their uncertainties. 

\begin{deluxetable*}{lcccc}
\tabletypesize{\tiny}
\tablecaption{Summary of stellar parameters. \label{tab:stellarparam}}
\tablehead{\colhead{~~~Parameter}&  \colhead{Description}&
\colhead{TOI-3984 A}&
\colhead{TOI-5293 A}&
\colhead{Reference}}
\startdata
\multicolumn{4}{l}{\hspace{-0.2cm} Main identifiers:}  \\
~~~TIC & \nodata  & 20182780 & 250111245 & TIC \\
~~~Gaia DR3 & \nodata & 1291955578869575552 & 2640121486388076032 & Gaia DR3 \\
\hline
\multicolumn{4}{l}{\hspace{-0.2cm} Coordinates, proper motion, distance, maximum extinction, and spectral type:} \\
~~~$\alpha_{\mathrm{J2016}}$ &  Right Ascension (RA) & 15:05:20.91 & 23:43:18.88 & Gaia DR3 \\
~~~$\delta_{\mathrm{J2016}}$ &  Declination (Dec) & 36:47:13.14 & -02:02:42.33 & Gaia DR3 \\
~~~$\mu_{\alpha}$ &  Proper motion (RA, mas yr$^{-1}$) & $-48.95 \pm 0.02$ & $-17.13 \pm 0.03$ & Gaia DR3 \\
~~~$\mu_{\delta}$ &  Proper motion (Dec, mas yr$^{-1}$) & $42.65 \pm 0.02$ & $0.49 \pm 0.02$ & Gaia DR3  \\
~~~$l$ &  Galactic longitude & 60.26769 & $87.00510$ & Gaia DR3 \\
~~~$b$ &  Galactic latitude & $60.17839$ & $-60.07625$ & Gaia DR3  \\
~~~$d$ &  Geometric distance (pc)  & $108.4_{-0.2}^{+0.3}$ & $160.8 \pm 0.6$ & Bailer-Jones\\
~~~\(A_{V,max}\) & Maximum visual extinction & $0.04$ & $0.06$ & Green\\
~~~Spectral type & \nodata & M$4\pm0.5$ & M$3\pm1$ & LAMOST/This work$^a$ \\
\hline
\multicolumn{4}{l}{\hspace{-0.2cm} Broadband photometric magnitudes:}  \\
~~~$B$ & Johnson $B$ mag & \nodata & $17.4 \pm 0.2$ & APASS\\
~~~$V$ & Johnson $V$ mag & \nodata & $15.9 \pm 0.2$ & APASS\\
~~~$g'$ & Pan-STARRS1 $g'$ mag & $16.35 \pm 0.02$ & $16.496 \pm 0.007$ & PS1\\
~~~$r'$ & Pan-STARRS1 $r'$ mag & $15.18 \pm 0.01$ & $15.29 \pm 0.01$ & PS1\\
~~~$i'$ & Pan-STARRS1 $i'$ mag & $13.96 \pm 0.01$ & $14.310 \pm 0.008$ & PS1\\
~~~$z'$ & Pan-STARRS1 $z'$ mag & $13.41 \pm 0.02$ & $13.87 \pm 0.01$ & PS1\\
~~~$y'$ & Pan-STARRS1 $y'$ mag & $13.15 \pm 0.02$ & $13.657 \pm 0.008$ & PS1\\
~~~$J$ & $J$ mag & $11.93 \pm 0.02$ & $12.47 \pm 0.03$ & 2MASS\\
~~~$H$ & $H$ mag & $11.32 \pm 0.02$ & $11.86 \pm 0.03$ & 2MASS\\
~~~$K_s$ & $K_s$ mag & $11.05 \pm 0.02$ & $11.64 \pm 0.04$ & 2MASS\\
~~~W1 & WISE1 mag & $10.94 \pm 0.02$ & $11.37 \pm 0.02$ & WISE\\
~~~W2 & WISE2 mag & $10.85 \pm 0.02$ & $11.30 \pm 0.02$ & WISE\\
~~~W3 & WISE3 mag & $10.69 \pm 0.06$ & $11.2 \pm 0.2$   & WISE\\
\hline
\multicolumn{4}{l}{\hspace{-0.2cm} Spectroscopic parameters$^b$:}\\
~~~$T_{e}$ &  Effective temperature (K) & $3476\pm88$ & $3586\pm88$ & This work\\
~~~$\log g_\star$ & Surface gravity (cgs) & $4.81\pm0.05$ & $4.77\pm0.05$ & This work\\
~~~$\mathrm{[Fe/H]}$ &  Metallicity (dex) & $0.18\pm0.12$ & $-0.03\pm0.12$  & This work\\
~~~$v\sin i_{\star}$ & Rotational broadening (km s$^{-1}$) & $<2$ & $<2$ & This work\\
\hline
\multicolumn{4}{l}{\hspace{-0.2cm} Model-dependent parameters from a stellar SED and isochrone fit$^c$:}\\
~~~$M_\star$ &  Mass ($M_{\odot}$) & $0.49 \pm 0.02$ & $0.54 \pm 0.02$ & This work\\
~~~$R_\star$ &  Radius ($R_{\odot}$) & $0.47 \pm 0.01$ & $0.52_{-0.01}^{+0.02}$ & This work\\
~~~$\rho_\star$ &  Density ($\mathrm{g~cm^{-3}}$) & $6.9 \pm 0.5$ & $5.4 \pm 0.4$ & This work\\
~~~$A_v$ & Visual extinction (mag) & $0.02 \pm 0.01$ & $0.04 \pm 0.03$ & This work\\
\hline
\multicolumn{4}{l}{\hspace{-0.2cm} Other stellar parameters:}\\
~~~Rotation Period & Days & $50.0_{-2.7}^{+2.8}$ & $20.6_{-0.4}^{+0.3}$ & This work\\
~~~Age$^d$ & Age (Gyrs) & $0.7-5.1$ & $0.7-5.1$ & This work\\
~~~$RV$ & Systemic RV (km s$^{-1}$) & $-5.77\pm0.06$ & $8.0\pm0.1$ & This work\\
~~~$U, V, W$ &  Barycentric Galactic velocities (km s$^{-1}$) &  $-33.01 \pm 0.07,~ -5.86 \pm 0.03,~ 5.65 \pm 0.06$ & $11.56 \pm 0.06,~ 9.63 \pm 0.06,~ -3.32 \pm 0.09$ & This work\\
~~~$(U, V, W)_{\mathrm{LSR}}$ &  Galactic velocities w.r.t. LSR$^e$ (km s$^{-1}$) &  $-21.9 \pm 0.8,~ 6.4 \pm 0.5,~ 12.9 \pm 0.4$ & $22.7 \pm 0.8,~ 21.9 \pm 0.5,~ 3.9 \pm 0.4$ & This work\\
\enddata
\tablerefs{TIC \citep{Stassun2019}, Gaia DR3 \citep{GaiaCollaboration2022}, Bailer-Jones \citep{Bailer-Jones2021}, Green \citep{Green2019}, LAMOST \citep{Zhong2019}, APASS \citep{Henden2018}, PS1 \citep{Chambers2016}, 2MASS \citep{Cutri2003}, WISE \citep{Wright2010}}
\tablenotetext{a}{Spectral type for TOI-3984 A is from LAMOST DR8. Spectral type for TOI-5293 A is derived with \texttt{PyHammer}.}
\tablenotetext{b}{Derived with the \texttt{HPF-SpecMatch} package.}
\tablenotetext{c}{Derived with the \texttt{EXOFASTv2} package using MIST isochrones.}
\tablenotetext{d}{Ages based on the rotation period and corresponding age range from \cite{Newton2016}.}
\tablenotetext{e}{Calculated using the solar velocities from \cite{Schoenrich2010}.}
\end{deluxetable*}

\subsection{Spectral classification}
\subsubsection{LAMOST}
The Large Sky Area Multi-Object Fibre Spectroscopic Telescope (LAMOST) is a 4 m telescope equipped with 4000 fibers distributed over a 5\degr\ FOV that is capable of acquiring spectra in the optical band (3700-9000\AA) at \(R\approx1800\) \citep{Cui2012}. TOI-3984 A was observed as part of its spectroscopic survey of the Galaxy \citep{Deng2012,Yuan2015,Xiang2017} while TOI-5293 A has not been observed. The data used in this work are from the public DR8v2.0\footnote{\url{http://dr8.lamost.org/}} release \citep{Wang2022}. 

The LAMOST stellar classification pipeline uses stellar templates to identify molecular absorption features (e.g., CaH, TiO) typical for M-type stars \citep{Lepine2007} and has been shown to report the subclass of an M dwarf with an accuracy of $\pm0.5$ subtypes \citep[][]{Zhong2015}. A successful classification requires that a target have (i) a mean S/N$>5$, (ii) a best-matching template that is an M type, and (iii) molecular band indices that are located in the M-type stellar regime identified by \cite{Zhong2019} ($0<\mathrm{TiO5}< 1.2$ and $0.6<\mathrm{CaH2+CaH3}< 2.4$). LAMOST classifies TOI-3984 A as an M$4\pm0.5$ dwarf, which agrees with the derived parameters in Section \ref{sec:specmatch}.

\subsubsection{LRS2}
The second generation Low Resolution Spectrograph \citep[LRS2;][]{Chonis2014,Chonis2016} is a low-resolution, optical integral-field unit spectrograph on the HET. LRS2 
has broad wavelength coverage spread between two fiber-fed, dual-channel spectrographs which simultaneously observe independent fields separated by 100\arcsec. The blue spectrograph pair (LRS2-B) covers $364\mathrm{nm} \le \lambda \le 467\mathrm{nm}$ and $454\mathrm{nm} \le \lambda \le 700\mathrm{nm}$ with $R\sim2500$ and $R\sim1400$, respectively. The red spectrograph pair (LRS2-R) covers $643\mathrm{nm} \le \lambda \le 845\mathrm{nm}$ and  $823\mathrm{nm} \le \lambda \le 1056\mathrm{nm}$ at $R\sim2500$. TOI-5293 A was observed with LRS2-R on 2022 November 01 and LRS2-B on 2022 December 11 using exposure times of 1800s. 

The LRS2 data were processed using the automated pipeline, \texttt{Panacea}\footnote{\href{https://github.com/grzeimann/Panacea}{https://github.com/grzeimann/Panacea~\faGithub}}, and the package \texttt{LRS2Multi}\footnote{\href{https://github.com/grzeimann/LRS2Multi}{https://github.com/grzeimann/LRS2Multi~\faGithub}}. We follow the methodology outlined in \cite{Kanodia2023}. Briefly, \texttt{Panacea} performs bias-correction, wavelength calibration, fiber extraction, and an initial flux calibration while \texttt{LRS2Multi} extracts spectra from \texttt{Panacea} products. The stellar spectra are extracted from a 3.5\arcsec\ aperture centered on TOI-5293 A using sky-subtracted frames. The response correction was derived using calibrated standard stars observed between 2021 May through 2022 May while the telluric correction was constructed using telluric standard stars. 

We use \texttt{PyHammer} \citep{Kesseli2017,Roulston2020} to estimate the spectral type of TOI-5293. \texttt{PyHammer} assigns a spectral type by measuring spectral indices for various atomic and molecular lines and comparing the values to those measured from observed stellar templates. The M dwarf templates used by \texttt{PyHammer} are from the MaNGA Stellar Library \citep[$362.2-1035.4~nm$ at $R\sim1800$;][]{Yan2019}. \texttt{PyHammer} selects the best-matching spectral type by minimizing the $\chi^2$ difference between spectral indices of the stellar template and observed star. We estimate a spectral type of M$3\pm1$ from the LRS2 data, which agrees with the derived parameters in Section \ref{sec:specmatch}.

\subsection{Spectral energy distribution fitting}
We modeled the spectral energy distribution (SED, see more details in Appendix \ref{app:sed}) for each target using the \texttt{EXOFASTv2} analysis package \citep{Eastman2019} to derive model-dependent stellar parameters. \texttt{EXOFASTv2} calculates the bolometric corrections for the SED fit by linearly interpolating the precomputed bolometric corrections\footnote{\url{http://waps.cfa.harvard.edu/MIST/model_grids.html\#bolometric}} in \(\log g_\star\), \(\mathrm{T_{e}}\), [Fe/H], and \(A_V\) from the MIST model grids \citep{Dotter2016,Choi2016}. Table \ref{tab:stellarparam} contains the stellar priors and derived stellar parameters with their uncertainties. The model-dependent mass and radius are (i) \(0.49 \pm 0.02~\mathrm{M_{\odot}}\) and \(0.47 \pm 0.01~\mathrm{R_{\odot}}\) for TOI-3984 A and (ii) \(0.54 \pm 0.02~\mathrm{M_{\odot}}\) and \(0.52_{-0.01}^{+0.02}~\mathrm{R_{\odot}}\) for TOI-5293 A.

\subsection{Stellar rotation period} \label{sec:prot}
We search the ZTF photometry for a rotation period of any activity-induced photometric modulations in TOI-3984 A and TOI-5293 A using the \texttt{GLS}\footnote{\href{https://github.com/mzechmeister/GLS}{https://github.com/mzechmeister/GLS~\faGithub}} package. We only consider significant peaks in the periodogram calculated with \texttt{GLS} where the false alarm probability (FAP), as calculated following \cite{Zechmeister2009}, is below a threshold of $0.1\%$. Data within transits were excised using the duration and ephemeris from the QLP. Significant peaks (FAP$<0.1\%$) of $\sim52$ were found in both the $zr$ and $zg$ photometry of TOI-3984 A while a significant peak of $\sim21$ days was seen only in the $zr$ photometry of TOI-5293 A.

The measured rotation periods are \(50.0_{-2.7}^{+2.8}\) days for TOI-3984 A and \(20.6_{-0.4}^{+0.3}\) days for TOI-5293 A (see a detailed description of the measurement in Appendix \ref{app:prot}). These intermediate rotation periods suggest both TOI-3984 A and TOI-5293 A most likely have ages between $0.7-5.1$ Gyr if we adopt the classification scheme of \cite{Newton2016}. This range is consistent with the estimated rotation periods of $\sim3.1$ Gyr for TOI-3293 A and $\sim1.3$ Gyr for TOI-5293 A using the rotation period and age relationship from \cite{Engle2018} for M2.5-M6 dwarfs. The age estimates from the model-dependent SED fit of $7.9_{-5.0}^{+4.1}$ Gyr and $7.7_{-4.9}^{+4.1}$ Gyr for TOI-3984 A and TOI-5293 A, respectively, are also consistent with the age range from \cite{Newton2016}. 

\subsection{Galactic kinematics}
The \textit{UVW} velocities are derived with \texttt{galpy} \citep{Bovy2015} and provided with respect to the local standard of rest from \cite{Schoenrich2010} using the Gaia DR3 astrometry and the systemic velocity derived from HPF. The values in Table \ref{tab:stellarparam} are in a right-handed coordinate system \citep{Johnson1987} where \textit{UVW} are positive in the directions of the Galactic center, Galactic rotation, and the north Galactic pole, respectively. The BANYAN \(\Sigma\) algorithm \citep{Gagne2018}, which uses sky positions, proper motions, parallax, and RVs to constrain cluster membership probabilities, classifies both TOI-3984 A and TOI-5293 A as field stars having no membership with known nearby young associations. Both TOI-3984 A and TOI-5293 A are also classified as a member of the thin disk ($\mathrm{P_{Thick}}/\mathrm{P_{Thin}}<0.02$) following the kinematic selection criteria from \cite{Bensby2003}. 

\section{Photometric and RV modeling} \label{sec:modelfit}
We use the \texttt{juliet} analysis package \citep{Espinoza2019} to jointly model the RVs with a standard Keplerian RV curve generated from the \texttt{radvel} \citep{Fulton2018} package and the light curves with a transit model generated from the \texttt{batman} package \citep{Kreidberg2015}. \texttt{juliet} performs the parameter estimation using \texttt{dynesty} \citep{Speagle2020}. The photometric model adopts a quadratic limb-darkening law where the coefficients are sampled from uniform priors following the parameterization in \cite{Kipping2013b}. For the long-cadence TESS photometry, the transit model utilizes the supersampling option in \texttt{batman} with exposure times of 30 minutes and a supersampling factor of 30. 

The TESS photometric model includes a Gaussian process noise model identical to that described in Section \ref{sec:prot} to account for correlated noise. We do not expect the period of the Gaussian process model to reflect the rotation periods measured in Section \ref{sec:prot} because the TESS PDCSAP and \texttt{eleanor} \texttt{CORR\_FLUX} are generated using algorithms known to attenuate long-period ($>10$ days) signals \citep[e.g.,][]{VanCleve2016,Holcomb2022}. 

A dilution factor, \(D\), is included in the TESS photometric model to account for dilution from nearby stars in the large apertures (see Figures \ref{fig:3984apertures} and \ref{fig:5293apertures}). We assume the higher spatial resolution and smaller photometric apertures used for the ground-based photometry result in negligible contamination from neighboring stars, such that the dilution term is fixed to unity for all ground-based transits. The fit uses a uniform prior on the TESS dilution term of $0-2$ to account for potential over-compensation of the dilution term that can occasionally occur with short-cadence PDCSAP TESS light curves in crowded fields \citep[e.g.,][]{Burt2020}. Both the photometric and RV models include a white-noise model parameterized as a jitter term that is added in quadrature to the uncertainty of each data set.

Tables \ref{tab:3984par} and \ref{tab:5293par} provide the priors used for the fit along with the inferred system parameters and the confidence intervals ($16\mathrm{th}-84\mathrm{th}$ percentile) for TOI-3984 A and TOI-5293 A, respectively. The fits suggest a significant detection of the RV orbit for each system and we investigate this by comparing the Bayesian Information Criterion \citep[BIC;][]{Schwarz1978,Liddle2007} and the Akaike Information Criterion \citep[AIC;][]{Akaike1974}. We compute the AIC and BIC of the best fit RV curve (9 free parameters for the RV curve: 5 orbital elements and one offset and jitter per instrument) and a flat line (4 free parameters: one offset and jitter per instrument) to determine the significance of the detection. We use the likelihood function ($\mathcal{L}$) employed by \texttt{juliet} \citep[Equation 4 in][]{Espinoza2019} for this comparison. For TOI-3984 A, $\Delta \ln \mathcal{L}=13.5$, $\Delta\mathrm{AIC}=17.1$, and $\Delta\mathrm{BIC}=8.4$ between a flat line and Keplerian model, which provide support for the existence of a Keplerian orbit in the data. Similarly, for TOI-5293 A, $\Delta \ln \mathcal{L}=46.3$, $\Delta\mathrm{AIC}=82.7$, and $\Delta\mathrm{BIC}=78.2$ provide very strong evidence for a Keplerian orbit in the data.

Figures \ref{fig:3984tess}--\ref{fig:3984groundphot}, \ref{fig:3984rv}, and \ref{fig:5293rv} display the model posteriors for each system. TOI-3984 A b is a sub-Saturn ($M_{p}=0.14\pm0.03~\mathrm{M_J}$ and $R_{p}=0.71\pm0.02~\mathrm{R_J}$) on a nearly circular orbit with a period of \(4.353326 \pm 0.000005\) days. TOI-5293 A b is a hot Jupiter ($M_{p}=0.54\pm0.07~\mathrm{M_J}$ and $R_{p}=1.06\pm0.04~\mathrm{R_J}$) on a nearly circular orbit with a period of \(2.930289 \pm 0.000004\) days. 

We estimate the timescales for circularization using the formalism of \cite{Jackson2008}, assuming that the tidal quality factor of each planet spans $Q_p=10^5-10^8$, based on the observed range from \cite{Bonomo2017}, and the tidal quality factor of each host star is \(Q_{\star}=10^7\), based on the modeling of \cite{Gallet2017}. The timescale for circularization spans $0.1-10$ Gyr when using the orbit parameters derived from the joint fit. With no precise age or $Q_p$ constraint, these systems may be able to retain a small but non-zero eccentricity. Existing data are only sufficient to confirm these are low-eccentricity planetary systems with $3\sigma$ upper limits of $e<0.23$ for TOI-3984 A b and $e<0.38$ for TOI-5293 A b. Higher-precision photometric observations of the occultation \citep[e.g.,][]{Alonso2018} would be most sensitive to measuring any non-zero eccentricity.

\begin{deluxetable*}{llccccccc}
\tabletypesize{\fontsize{4}{5}\selectfont}
\rotate
\tablecaption{System parameters for TOI-3984 A \label{tab:3984par}}
\tablehead{\colhead{~~~Parameter} &
\colhead{Units} &
\colhead{Prior} &
\multicolumn{6}{c}{Value} 
}
\startdata
\noalign{\vskip 1.5ex} Ground-based photometry parameters & & & RBO (07-19) & RBO (08-14) & Soka & RBO (02-04) & ARCTIC (03-23) & ARCTIC (05-10)\\ \noalign{\vskip .8ex}
~~~Linear limb-darkening coefficient$^a$ & $q_1$ & $\mathcal{U}(0,1)$ & $0.2 \pm 0.1$ & $0.2 \pm 0.1$ & $0.09_{-0.02}^{+0.03}$ & $0.2 \pm 0.1$ & $0.09_{-0.02}^{+0.03}$ &  $0.05_{-0.02}^{+0.03}$ \\
~~~Quadratic limb-darkening coefficient$^a$ & $q_2$ & $\mathcal{U}(0,1)$ & $0.5 \pm 0.2$ & $0.5 \pm 0.2$ & $0.8_{-0.2}^{+0.1}$ & $0.5 \pm 0.2$ & $0.8_{-0.2}^{+0.1}$ & $0.6 \pm 0.2$ \\
~~~Photometric jitter & $\sigma_{phot}$ (ppm) & $\mathcal{J}(10^{-6},10^{4})$ & $0.0001_{-0.0001}^{+0.0261}$ & $0.3_{-0.3}^{+20.4}$ & $0.0001_{-0.0001}^{+0.0045}$ & $0.02_{-0.02}^{+2.35}$ & $0.3_{-0.3}^{+24.1}$ & $989_{-9}^{+7}$\\
\hline
\noalign{\vskip 1.5ex} TESS photometry parameters & & & \multicolumn{3}{c}{Long-cadence} & \multicolumn{3}{c}{Short-cadence} \\ 
\noalign{\vskip .8ex}
~~~Linear limb-darkening coefficient$^a$ & $q_1$ & $\mathcal{U}(0,1)$ & \multicolumn{3}{c}{$0.24_{-0.09}^{+0.13}$}  & \multicolumn{3}{c}{$0.24_{-0.09}^{+0.13}$}  \\
~~~Quadratic limb-darkening coefficient$^a$ & $q_2$ & $\mathcal{U}(0,1)$ & \multicolumn{3}{c}{$0.6_{-0.3}^{+0.2}$} & \multicolumn{3}{c}{$0.6_{-0.3}^{+0.2}$}  \\
~~~Photometric jitter & $\sigma_{phot}$ (ppm) & $\mathcal{J}(10^{-6},10^{3})$ & \multicolumn{3}{c}{$0.01_{-0.01}^{+2.56}$} & \multicolumn{3}{c}{$10_{-10}^{+80}$} \\
~~~Dilution factor & $D$ & $\mathcal{U}(0,2)$ & \multicolumn{3}{c}{$0.89_{-0.03}^{+0.04}$} & \multicolumn{3}{c}{$0.87_{-0.03}^{+0.04}$}  \\
\hline
\sidehead{TESS Gaussian process hyperparameter:}
~~~$B$ & Amplitude (ppm) & $\mathcal{J}(10^{-4},10^{12})$ & \multicolumn{3}{c}{$37_{-7}^{+9}$} & \multicolumn{3}{c}{$10_{-4}^{+779}$} \\
~~~$C$ & Additive factor  & $\mathcal{J}(10^{-3},10^3)$ &  \multicolumn{3}{c}{$0.02_{-0.01}^{+0.26}$} & \multicolumn{3}{c}{$1.4_{-1.3}^{+244.2}$} \\
~~~$L$ & Length scale (days)  & $\mathcal{J}(10^{-3},10^3)$ & \multicolumn{3}{c}{$1.7_{-0.4}^{+0.6}$} & \multicolumn{3}{c}{$1.8_{-0.8}^{+133.6}$}  \\
~~~$P_{GP}$ & Period (days)  & $\mathcal{J}(1.0,100)$ & \multicolumn{3}{c}{$2.7_{-0.5}^{+1.2}$} & \multicolumn{3}{c}{$2.7_{-0.5}^{+1.2}$} \\
\hline
\noalign{\vskip 1.5ex} RV parameters & & & \multicolumn{3}{c}{HPF} & \multicolumn{3}{c}{NEID} \\ 
~~~Systemic velocity & $\gamma~\mathrm{(m~s^{-1})}$ & $\mathcal{U}(-10^3,10^3)$ & \multicolumn{3}{c}{$2.4_{-3.8}^{+4.0}$} & \multicolumn{3}{c}{$-57.7_{-10.4}^{+10.7}$} \\
~~~RV Jitter & $\sigma_{RV}~\mathrm{(m~s^{-1})}$ & $\mathcal{J}(10^{-3},10^3)$ & \multicolumn{3}{c}{$3.1_{-2.8}^{+8.0}$} & \multicolumn{3}{c}{$29.0_{-9.8}^{+14.0}$}\\
\hline
\sidehead{Orbital parameters:}
~~~Orbital period & $P$ (days)  & $\mathcal{N}(4.35,0.1)$ & \multicolumn{6}{c}{$4.353326 \pm 0.000005$}\\
~~~Time of mid-transit & $T_C$ (BJD\textsubscript{TDB}) & $\mathcal{N}(2459715.02,0.1)$ & \multicolumn{6}{c}{$2459715.02268 \pm 0.00009$}\\
~~~$\sqrt{e}\cos\omega$ &  & $\mathcal{U}(-1,1)$ & \multicolumn{6}{c}{$-0.07_{-0.17}^{+0.23}$}\\
~~~$\sqrt{e}\sin\omega$ &  & $\mathcal{U}(-1,1)$ & \multicolumn{6}{c}{$-0.05_{-0.10}^{+0.09}$}\\
~~~Semi-amplitude velocity & $K~\mathrm{(m~s^{-1})}$  & $\mathcal{U}(0,10^3)$ &   \multicolumn{6}{c}{$27.7_{-5.0}^{+5.4}$}\\
~~~Scaled radius & $R_{p}/R_{\star}$  & $\mathcal{U}(0,1)$ & \multicolumn{6}{c}{$0.1558_{-0.0008}^{+0.0009}$}\\
~~~Impact parameter & $b$ & $\mathcal{U}(0,1)$ & \multicolumn{6}{c}{$0.18_{-0.07}^{+0.06}$}\\
~~~Scaled semi-major axis & $a/R_{\star}$  & $\mathcal{J}(1,100)$ & \multicolumn{6}{c}{$19.1_{-0.4}^{+0.5}$}\\
\hline
\sidehead{Derived parameters:}
~~~Eccentricity & $e$  & \nodata & \multicolumn{6}{c}{$0.04_{-0.03}^{+0.05}$, $3\sigma<0.23$}\\
~~~Argument of periastron & $\omega$ (degrees)  & \nodata& \multicolumn{6}{c}{$-54_{-108}^{+189}$}\\
~~~Orbital inclination & $i$ (degrees) & \nodata & \multicolumn{6}{c}{$89.5 \pm 0.2$}\\
~~~Transit duration & $T_{14}$ (hours) & \nodata & \multicolumn{6}{c}{$2.01 \pm 0.01$}\\
~~~Mass & $M_{p}$  ($\mathrm{M_{J}}$ / $\mathrm{M_{\oplus}}$)  & \nodata &  \multicolumn{6}{c}{$0.14\pm0.03$ / $44.0_{-8.0}^{+8.7}$}\\
~~~Radius & $R_{p}$  ($\mathrm{R_{J}}$ / $\mathrm{R_{\oplus}}$)  & \nodata &  \multicolumn{6}{c}{$0.71\pm0.02$ / $7.9\pm0.24$}\\
~~~Surface gravity & $\log g_{p}$  (cgs)  & \nodata &  \multicolumn{6}{c}{$2.84\pm0.08$}\\
~~~Density & $\rho_{p}$  ($\mathrm{g~cm}^{-3}$)  & \nodata &  \multicolumn{6}{c}{$0.49_{-0.10}^{+0.11}$}\\
~~~Semi-major axis & $a$ (au)  & \nodata & \multicolumn{6}{c}{$0.041_{-0.001}^{+0.002}$}\\
~~~Average Incident flux & $\langle F \rangle$ ($\mathrm{10^8\ erg~s^{-1}~cm^{-2}}$ / $S_\oplus$) & \nodata & \multicolumn{6}{c}{$0.23_{-0.02}^{+0.03}$ / $16.7\pm1.8$}\\
~~~Equilibrium temperature\(^{b}\) & $T_{eq}$ (K) & \nodata & \multicolumn{6}{c}{$563\pm15$}\\
\enddata
\tablenotetext{a}{Using the $q1$ and $q2$ parameterization from \cite{Kipping2013b}.}
\tablenotetext{b}{The planet is assumed to be a black body and we ignore heat redistribution.}
\end{deluxetable*}

\begin{deluxetable*}{llcccccc}
\tabletypesize{\tiny}
\tablecaption{System parameters for TOI-5293 A \label{tab:5293par}}
\tablehead{\colhead{~~~Parameter} &
\colhead{Units} &
\colhead{Prior} &
\multicolumn{5}{c}{Value} 
}
\startdata
\noalign{\vskip 1.5ex} Ground-based photometry parameters & & & RBO (07-28) & RBO (09-13) & LCRO & TMMT (10-24) & TMMT (10-27) \\ \noalign{\vskip .8ex}
~~~Linear limb-darkening coefficient$^a$ & $q_1$ & $\mathcal{U}(0,1)$ & $0.4_{-0.2}^{+0.3}$ & $0.4_{-0.2}^{+0.3}$ & $0.5 \pm 0.2$ & $0.3_{-0.2}^{+0.3}$ & $0.3_{-0.2}^{+0.3}$ \\
~~~Quadratic limb-darkening coefficient$^a$ & $q_2$ & $\mathcal{U}(0,1)$ & $0.3_{-0.2}^{+0.3}$ & $0.3_{-0.2}^{+0.3}$ & $0.5 \pm 0.3$ & $0.3 \pm 0.2$ & $0.3 \pm 0.2$ \\
~~~Photometric jitter & $\sigma_{phot}$ (ppm) & $\mathcal{J}(10^{-6},10^{3})$ & $0.007_{-0.007}^{+2.846}$ & $0.04_{-0.04}^{+12.96}$ & $0.1_{-0.1}^{+34.6}$ & $0.05_{-0.05}^{+18.11}$ & $0.2_{-0.2}^{+40.7}$ \\
\hline
\sidehead{TESS photometry parameters:}
~~~Linear limb-darkening coefficient$^a$ & $q_1$ & $\mathcal{U}(0,1)$ & \multicolumn{5}{c}{$0.4 \pm 0.2$}  \\
~~~Quadratic limb-darkening coefficient$^a$ & $q_2$ & $\mathcal{U}(0,1)$ & \multicolumn{5}{c}{$0.3 \pm 0.2$} \\
~~~Photometric jitter & $\sigma_{phot}$ (ppm) & $\mathcal{J}(10^{-6},10^{3})$ & \multicolumn{5}{c}{$0.2_{-0.2}^{+37.5}$} \\
~~~Dilution factor & $D$ & $\mathcal{U}(0,2)$ & \multicolumn{5}{c}{$0.94_{-0.03}^{+0.04}$} \\
\hline
\sidehead{TESS Gaussian process hyperparameter:}
~~~$B$ & Amplitude & $\mathcal{J}(10^{-10},10^{6})$ & \multicolumn{5}{c}{$0.0010_{-0.0005}^{+0.0015}$} \\
~~~$C$ & Additive factor  & $\mathcal{J}(10^{-3},10^3)$ & \multicolumn{5}{c}{$0.2_{-0.2}^{+5.5}$} \\
~~~$L$ & Length scale (days)  & $\mathcal{J}(10^{-3},10^3)$ & \multicolumn{5}{c}{$40_{-22}^{+63}$} \\
~~~$P_{GP}$ & Period (days)  & $\mathcal{J}(1.0,100)$ & \multicolumn{5}{c}{$16_{-5}^{+10}$} \\
\hline
\sidehead{RV parameters:}
~~~Systemic velocity & $\gamma~\mathrm{(m~s^{-1})}$ & $\mathcal{U}(-10^3,10^3)$ & \multicolumn{5}{c}{$21.3_{-11.4}^{+10.8}$} \\
~~~RV Jitter & $\sigma_{RV}~\mathrm{(m~s^{-1})}$ & $\mathcal{J}(10^{-3},10^3)$ & \multicolumn{5}{c}{$27.4_{-9.8}^{+12.3}$}\\
\hline
\sidehead{Orbital parameters:}
~~~Orbital period & $P$ (days)  & $\mathcal{N}(2.9,0.1)$ & \multicolumn{5}{c}{$2.930289 \pm 0.000004$}\\
~~~Time of mid-transit & $T_0$ (BJD\textsubscript{TDB}) & $\mathcal{N}(2459448.9,0.1)$ & \multicolumn{5}{c}{$2459448.9148 \pm 0.0004$}\\
~~~$\sqrt{e}\cos\omega_\star$ &  & $\mathcal{U}(-1,1)$ & \multicolumn{5}{c}{$-0.07_{-0.16}^{+0.17}$}\\
~~~$\sqrt{e}\sin\omega_\star$ &  & $\mathcal{U}(-1,1)$ & \multicolumn{5}{c}{$-0.17\pm0.22$}\\
~~~Semi-amplitude velocity & $K~\mathrm{(m~s^{-1})}$  & $\mathcal{U}(0,10^3)$ &   \multicolumn{5}{c}{$115.6\pm14.5$}\\
~~~Scaled radius & $R_{p}/R_{\star}$  & $\mathcal{U}(0,1)$ & \multicolumn{5}{c}{$0.210_{-0.004}^{+0.005}$}\\
~~~Impact parameter & $b$ & $\mathcal{U}(0,1)$ & \multicolumn{5}{c}{$0.32_{-0.14}^{+0.12}$}\\
~~~Scaled semi-major axis & $a/R_{\star}$  & $\mathcal{J}(1,100)$ & \multicolumn{5}{c}{$14.1_{-1.1}^{+1.6}$}\\
\hline
\sidehead{Derived parameters:}
~~~Eccentricity & $e$  & \nodata & \multicolumn{5}{c}{$0.08_{-0.06}^{+0.11}$, $3\sigma<0.38$}\\
~~~Argument of periastron & $\omega_\star$ (degrees)  & \nodata& \multicolumn{5}{c}{$-92_{-45}^{+161}$}\\
~~~Orbital inclination & $i$ (degrees) & \nodata & \multicolumn{5}{c}{$88.8_{-0.6}^{+0.5}$}\\
~~~Transit duration & $T_{14}$ (hours) & \nodata & \multicolumn{5}{c}{$1.94_{-0.04}^{+0.05}$}\\
~~~Mass & $M_{p}$  ($\mathrm{M_{J}}$ / $\mathrm{M_{\oplus}}$)  & \nodata &  \multicolumn{5}{c}{$0.54\pm0.07$ / $170.4_{-21.9}^{+21.8}$}\\
~~~Radius & $R_{p}$  ($\mathrm{R_{J}}$ / $\mathrm{R_{\oplus}}$)  & \nodata &  \multicolumn{5}{c}{$1.06\pm0.04$ / $11.9\pm0.4$}\\
~~~Surface gravity & $\log g_{p}$  (cgs)  & \nodata &  \multicolumn{5}{c}{$3.11_{-0.11}^{+0.12}$}\\
~~~Density & $\rho_{p}$  ($\mathrm{g~cm}^{-3}$)  & \nodata &  \multicolumn{5}{c}{$0.56\pm0.09$}\\
~~~Semi-major axis & $a$ (au)  & \nodata & \multicolumn{5}{c}{$0.034_{-0.003}^{+0.004}$}\\
~~~Average Incident flux & $\langle F \rangle$ ($\mathrm{10^8\ erg~s^{-1}~cm^{-2}}$ / $S_\oplus$) & \nodata & \multicolumn{5}{c}{$0.47_{-0.09}^{+0.12}$ / $34.6_{-6.2}^{+8.6}$}\\
~~~Equilibrium temperature\(^{b}\) & $T_{eq}$ (K) & \nodata & \multicolumn{5}{c}{$675_{-30}^{+42}$}\\
\enddata
\tablenotetext{a}{Using the $q1$ and $q2$ parameterization from \cite{Kipping2013b}.}
\tablenotetext{b}{The planet is assumed to be a black body and we ignore heat redistribution.}
\end{deluxetable*}

\section{Discussion}\label{sec:discussion}
\subsection{The dynamical implication of wide separation companions}
Hot Jupiters are difficult to form \textit{in-situ} at their observed separations from host stars \citep[e.g.,][]{Dawson2018,Anderson2020,Poon2021} either through gravitational instability \citep[e.g.,][]{Boss1997,Durisen2007} or core accretion \citep[e.g.,][]{Perri1974,Pollack1996,Chabrier2014}. These systems are hypothesized to have migrated from larger distances via the loss of angular momentum due to gravitational interactions with the circumstellar gas disk \citep[e.g.,][]{Goldreich1980,Lin1986,Lin1996,Ida2008,Baruteau2014} or with other massive companions \citep[e.g.,][]{Rasio1996,Weidenschilling1996,Wu2003,Fabrycky2007,Petrovich2015,Petrovich2015a}. Gas disk migration predominantly results in circular planetary orbits which are well aligned with the spin axis of the host star while high eccentricity tidal migration preferentially results in planets on misaligned and eccentric orbits that may have large-separation companions. 

We search for wide companions using a list of spatially resolved binary stars from an analysis of proper motions in Gaia EDR3 \citep{El-Badry2021}. Wide separation binary systems in \cite{El-Badry2021} are flagged as having either a main sequence or white dwarf companion using the location of the companion on the Gaia color-absolute magnitude diagram \citep{El-Badry2018b}. TOI-3984 A has a white dwarf companion, Gaia DR3 1291955574574621056 (TIC 1101522311), at a projected distance of 3.27\arcsec{} or a projected separation of 356 au. TOI-5293 A has an M dwarf companion, Gaia DR3 2640121482094497024 (TIC 2052711961), at a projected distance of 3.57\arcsec{} or a projected separation of 579 au. The companions have a negligible probability \citep[$<4\times10^{-8}$, see a detailed description in Appendix A of ][]{El-Badry2021} of being a false detection as a result of chance alignment with a background source having spurious parallax and proper motion measurements. The NEID and HPF spectra are not contaminated by either companion as they lie outside the HPF fiber \citep[$\sim1.7\arcsec$ on-sky;][]{Kanodia2018a} and the NEID HR fiber \citep[$\sim0.9\arcsec$ on-sky;][]{Schwab2016}.

We use the \texttt{phot\_bp\_rp\_excess\_factor} to determine if any of the measured blue or red Gaia photometry of either bound companion is blended or contaminated \citep{Evans2018, Riello2021}.  The value of \texttt{phot\_bp\_rp\_excess\_factor} in the Gaia DR3 archive is known to have a strong color dependence and the Gaia documentation suggests using a corrected excess factor that compares the reported value to the expected excess factor at a given color as derived from sources with good quality photometry \cite[see the discussion in Section 6 of][]{Riello2021}. We calculate a corrected \texttt{phot\_bp\_rp\_excess\_factor}\footnote{using Table 2 and Equation 6 from \cite{Riello2021}} of 0.20 and 0.15 for TOI-3984 B and TOI-5923 B, respectively. The deviation from zero reveals that both companions have some contamination in their Gaia colors. We adopt these colors as nominal values to qualitatively characterize the wide separation companions but note that photometry and spectroscopy of each target is required to provide robust companion parameters.

The Gaia General Stellar Parameterizer from Photometry algorithm \cite[GSP-Phot;][]{Andrae2022} confirms that (i) the companion to TOI-3984 A is most likely a white dwarf based on its location on the Gaia color-magnitude diagram and (ii) the companion to TOI-5293 A is a main sequence M dwarf. TOI-5293 B is reported to be a mid M dwarf with a radius of $R_\star=0.26_{-0.08}^{+0.10}~\mathrm{R_\odot}$ \citep{Creevey2022,Creevey2022a} with a $\log g=4.72_{-0.14}^{+0.16}$ and $\mathrm{T_{e}}=3041_{-41}^{+280}$ K derived by the Gaia DR3 Multiple Star Classifier\footnote{\url{https://gea.esac.esa.int/archive/documentation/GDR3/Data_analysis/chap_cu8par/sec_cu8par_apsis/ssec_cu8par_apsis_msc.html}} (MSC) analysis of Gaia BP/RP spectra.

Neither GSP-Phot nor MSC provide stellar parameters for TOI-3984 B and we use the \texttt{WD\_models}\footnote{\href{https://github.com/SihaoCheng/WD_models}{https://github.com/SihaoCheng/WD\_models~\faGithub}} package from \cite{Cheng2019} to derive a photometric age and mass. We assume the atmosphere is composed of hydrogen, due to the prevalence of DA white dwarfs in the 100 pc SDSS sample \citep[$\sim65\%$ of the sample;][]{Kilic2020,Kepler2021} and adopt the cooling models of \cite{Bedard2020}. The estimated mass for the white dwarf companion is $M_{WD}\sim0.75~\mathrm{M_\odot}$ with a cooling age of $\sim2.9$ Gyr (see Appendix \ref{app:wdloc}) and its progenitor star has a mass in the range $1.6-4.5~\mathrm{M_\odot}$ when using the MIST semi-empirical white dwarf initial-final mass relationship from \cite{Cummings2018}.  We note the cooling age is consistent with the age estimate of TOI-3984 A from the rotation period in Section \ref{sec:prot} ($0.7-5.1$ Gyr).

Various studies have been performed on the hot Jupiter population orbiting Sun-like stars to test the significance of multi-body interactions \citep[e.g.,][]{Wang2014a,Knutson2014a,Ngo2015,Ngo2016,Evans2018a,Ziegler2018b,Marzari2019,Fontanive2019,Belokurov2020,Hwang2020,Fontanive2021,Ziegler2021}. Hot Jupiters orbiting Sun-like stars have been reported to have a high wide binary fraction, relative to field stars. \cite{Ngo2016} measured that $47\pm7\%$ of systems with a hot Jupiter have resolved stellar companions between separations of $50 - 2000$ au while \cite{Fontanive2019} determined that $79.0^{+13.2}_{-14.7}\%$ of systems with a massive substellar object have a wide companion between 20-10000 au. \cite{Moe2021} reassessed these claims by accounting for biases and selection effects in the sample to conclude that wide separation companions ($50-2000$ au) to hosts of hot Jupiters do not enhance the formation of hot Jupiters at a statistically significant level and that the larger wide binary fraction of hot Jupiters is a result of suppressed hot Jupiter formation in close binaries. This non-enrichment in wide separation companions has been observed in the hot Jupiter population orbiting Sun-like stars with Gaia \citep{Hwang2020}.

The small population of hot Jupiters orbiting M dwarfs precludes an extensive study of stellar companions. Only two additional M dwarfs in the existing population of 10 transiting hot Jupiters ($P<10$ days, $R_p>8~\mathrm{R_\oplus}$), HATS-74A \citep{Jordan2022} and TOI-3714 \citep{Canas2022b}, are known to have wide separation companions. HATS-74A has an M dwarf companion at a separation of 238 au while TOI-3714 has a white dwarf companion, potentially on an eccentric orbit, at a projected separation of 302 au. 

The separation of the main sequence companions may be too large to strongly impact the formation of hot Jupiters, as some studies suggest only companions at separations of $\lesssim200$ au can impact the formation process \citep[e.g.,][]{Kraus2016,Moe2021,Ziegler2021}. The white dwarf progenitor in the TOI-3984 A system, however, would have been both more massive and much closer if we assume adiabatic mass loss \citep[see][]{Nordhaus2010,Nordhaus2013} and during the progenitor's main sequence lifetime, secular effects such as Kozai-Lidov cycles \citep{Kozai1962,Lidov1962,Naoz2016} could have induced high-eccentricity tidal migration of TOI-3984 A b that brought it to its observed location. 

Gaia DR3 is able to constrain the eccentricity of resolved wide binaries \citep[e.g.,][]{Tokovinin2020,Hwang2022} using the angle between the separation vector and the relative velocity vector ($v-r$ angle). \cite{Hwang2022} estimate the eccentricity of the wide binary sample identified by \cite{El-Badry2021} under the assumption that a wide companion has a random orbital orientation. The inferred eccentricities calculated by \cite{Hwang2022} are $e=0.64^{+0.18}_{-0.26}$ and $e=0.77_{-0.24}^{+0.16}$ for TOI-3984 B and TOI-5293 B, respectively. 

Similar to the white dwarf companion to TOI-3714, the high-eccentricity of TOI-3984 B is consistent with the scenario in which the progenitor star was initially on a smaller orbit, potentially inducing the migration of TOI-3984 A b, before later widening and becoming eccentric due to mass loss as it evolved into a white dwarf. While TOI-5293 B is also eccentric, the large projected separation ($579$ au) would result in a \(>5\) Gyr timescale for the Kozai-Lidov cycles \citep[Equation 7 from][]{Kiseleva1998} which would be too long to affect the migration of TOI-5293 A b, particularly given the most likely age of $0.7-5.1$ Gyr derived in Section \ref{sec:prot}.

\subsection{Constraints on additional planetary companions}
There are 15 confirmed planetary systems hosting a close-in gas giant ($P<100$ days and $M>0.6~\mathrm{M_J}$) with interior companions across all spectral types \cite[see Table A1 in][]{Sha2023}. For the transiting hot Jupiter sample ($P<10$ days), there are only 
6 confirmed gas giants with nearby companions: WASP-47 b \citep{Becker2015}, Kepler-730 b \citep{Zhu2018,Canas2019}, TOI-1130 c \citep{Huang2020a}, WASP-148 b \citep{Wang2022}, WASP-132 b \citep{Hord2022}, and TOI-2000 \citep{Sha2023}. This low planetary multiplicity rate for hot Jupiters orbiting Sun-like stars has been statistically confirmed with multiple ground and space-based transiting samples \citep[e.g.,][]{Steffen2012,Huang2016,Maciejewski2020,Hord2021,Wang2021a,Zhu2021}. A search for transit timing variations in the \textit{Kepler} sample by \cite{Wu2023} also determined that a minimum of $\sim13\%$ of hot Jupiters orbiting Sun-like stars have nearby companions and must have a quiescent formation history. The scarcity of short-period companions to hot Jupiters is consistent with the hypothesis that high-eccentricity migration is responsible for most of these systems, as the inward migration of a gas giant is a dynamically ``hot'' processes that would destabilize interior planets and leave an isolated gas giant \citep[e.g.,][]{Mustill2015,Dawson2018}.

We search for companions to TOI-3984 A and TOI-5923 using the available TESS photometry and HPF RVs. We used the transit least squares algorithm \citep[\texttt{TLS};][]{Hippke2019} on the photometry to search for additional transiting planets after subtracting the best-fitting transit model for each planet. For this search, we only searched for candidate signals (depths $>1$ ppm) between $1-13$ days. The maximum radius of a candidate signal was $\sim5\mathrm{~R_\oplus}$ for TOI-3984 A and $\sim7\mathrm{~R_\oplus}$ for TOI-5293 A. \texttt{TLS} identified no candidates where the test statistic was above the suggested threshold of 7 (corresponding to a false positive rate of $1\%$). The current TESS photometry is only sufficient to exclude the existence of additional short-period transiting gas giant companions. 

To further constrain the existence of non-transiting companions, we analyzed the residuals to the HPF RVs using \texttt{thejoker} \citep{Price-Whelan2017} to perform a rejection sampling analysis. Orbits were sampled using a log-uniform prior for the period (between 1 day and twice the HPF RV baseline), the Beta distribution from \cite{Kipping2013a} as a prior for the eccentricity, and a uniform prior for the argument of pericenter and the orbital phase. We considered a total of \(2^{28}\) samples using {\tt thejoker} and had a total acceptance rate of \(<0.07\%\) for both systems. The surviving samples place a $3\sigma$ upper limit on any coplanar ($\sin i\sim1$) companions of $M<2.4~\mathrm{M_J}$ ($K<\mathrm{150~m~s^{-1}}$) within 1 au ($P<520$ days) for TOI-3984 A\footnote{For comparison, $M<0.5~\mathrm{M_J}$ within 0.3 au} and $M<0.6~\mathrm{M_J}$ ($K<100\mathrm{~m~s^{-1}}$) within 0.3 au ($P<82$ days) for TOI-5293 A. The existing photometry and RVs reject the existence of nearby, coplanar massive planetary companions. Future observations with TESS and additional RVs are needed to provide robust constraints on the existence of small and low-mass transiting or non-transiting planetary companions.

\subsection{Comparison to the M dwarf planet population}
\begin{figure*}[!ht]
\epsscale{1.15}
\plotone{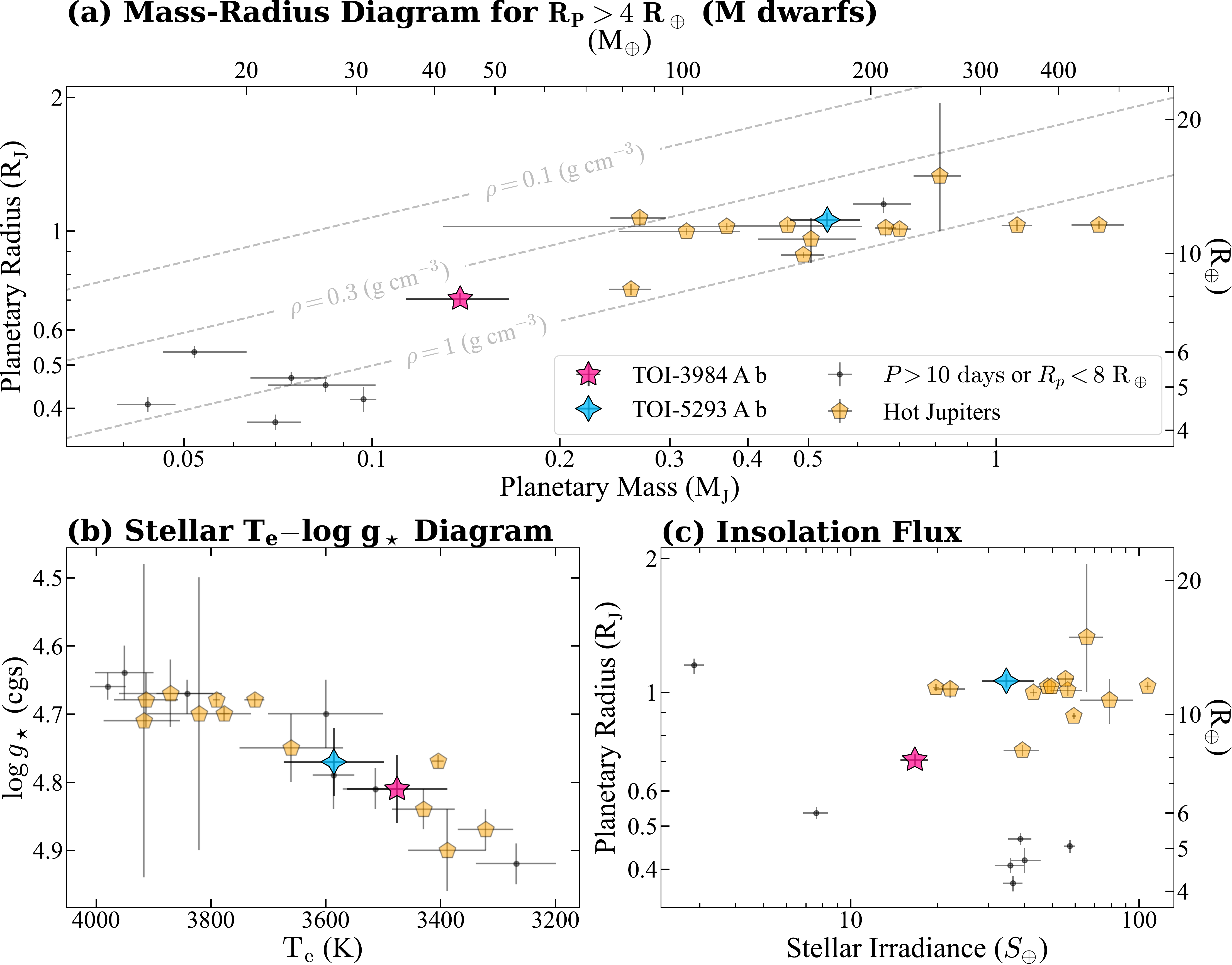}
\caption{\textbf{(a)} TOI-3984 A b (star) and TOI-5293 A b (diamond star) on the mass-radius diagram for transiting M dwarf exoplanets with mass measurements and $R_p>4~R_\oplus$. All previously known hot Jupiters ($P<10$ days and $R_P\ge8~\mathrm{R_\oplus}$) transiting M dwarfs are marked as pentagons. \textbf{(b)} TOI-3984 A and TOI-5293 A on an effective temperature $-$ surface gravity diagram. \textbf{(c)} the insolation flux and radius for these planets. The data were compiled from the \href{https://exoplanetarchive.ipac.caltech.edu/cgi-bin/TblView/nph-tblView?app=ExoTbls&config=PSCompPars}{NASA Exoplanet Archive} \citep{Akeson2013} on 2023 May 03.} 
\label{fig:planetprops}
\end{figure*}

A comparison of the TOI-3984 A and TOI-5293 A systems to the planetary mass--radius, stellar $T_{e} - \log g_\star$, and the period--insolation flux distributions of known M dwarf systems hosting planets with $R_p>4~\mathrm{R_\oplus}$ is shown in Figure \ref{fig:planetprops}. Among the population of short-period gas giants ($P<10$ days and $R_p\gtrsim8~\mathrm{R_\oplus}$) orbiting M dwarfs, TOI-3984 A b is a sub-Saturn with the smallest mass ($\sim0.47\mathrm{~M_{Saturn}}$) and radius ($\sim0.88\mathrm{~R_{Saturn}}$) while TOI-5293 A has a radius and mass consistent at $1\sigma$ to the median values of the existing M dwarf gas giant population ($1.02~\mathrm{R_J}$ and $0.50~\mathrm{M_J}$). The mid M dwarf nature of the host stars makes these two planets some of the coolest transiting M dwarf gas giants with an insolation flux of $S=16.7\pm1.8~S_{\oplus}$ for TOI-3984 A and $S=34.6_{-6.2}^{+8.6}~S_{\oplus}$ for TOI-5293 A, much lower than the typical transiting M dwarf gas giant (median value of $S\sim56~S_{\oplus}$). 

There are two additional hot Jupiters orbiting mid M dwarfs (M3-M5, $3400\mathrm{~K}<T_e<3600\mathrm{~K}$): HATS-71~b \citep{Bakos2020} and TOI-5205 b \citep{Kanodia2023} while the rest orbit early M dwarfs (M0-M2). The frequency of such hot Jupiters orbiting M dwarfs is theoretically predicted to be intrinsically low in the framework of core accretion because the low surface density of an M dwarf protoplanetary disk would impede growth of cores and the onset of runaway gas accretion \cite[e.g.,][]{Laughlin2004,Ida2005,Kennedy2008,Burn2021}. Constraints from radial velocity surveys \citep[e.g.,][]{Endl2006,Sabotta2021,Pinamonti2022} and photometric surveys \citep[e.g.,][]{Kovacs2013,Obermeier2016} have constrained the frequency of short-period gas giants orbiting M dwarfs to $\lesssim2\%$. Two independent searches for hot Jupiters orbiting M dwarfs in TESS by \cite{Gan2023} and \citep{Bryant2023} have yielded an occurrence rate of $0.27\pm0.09\%$ and $0.193\pm0.072\%$, respectively. 
 
 The measured occurrence rates for M dwarfs are smaller to recent values measured from TESS for Sun-like stars of $0.71\pm0.31\%$ \citep{Zhou2019a} and $0.98\pm0.36\%$ \citep[][]{Beleznay2022}. For comparison, the occurrence rate of hot Jupiters orbiting Sun-like stars has consistently been estimated to be $\sim1\%$ from various RV surveys such as the California Planet Survey \citep[$1.2\pm0.38\%$;][]{Wright2012}, the HARPS/CORALIE survey \citep[$0.9\pm0.4\%$;][]{Mayor2011}, and the Anglo-Australian Planet Search \citep[$0.84^{+0.7}_{-0.2}\%$;][]{Wittenmyer2020}. Results from the California Legacy Survey \citep{Zhu2022} suggest a higher occurrence rate of $2.8\pm0.8\%$, but \cite{Zhu2022} warn that the differences in the stellar population (such as binarity and metallicity) may result in a discrepancy from the canonical $\sim1\%$ occurrence rate for Sun-like stars. The current results for M dwarfs from \cite{Gan2023} and \cite{Bryant2023} suggest hot Jupiters are rarer as companions to M dwarfs, but these values are also within $1-3\sigma$ of most occurrence rates for Sun-like stars.
 
 The apparent dearth of hot Jupiters orbiting M dwarfs is expected if they predominantly form via core accretion. Giant planet formation must occur before depletion of the gas disk. The low mass and surface density of protoplanetary disks around M dwarfs \citep[e.g.,][]{Andrews2013,Mohanty2013,Stamatellos2015,Ansdell2017,Manara2018} stymie this process with a lower mass supply and longer timescales of planetesimal formation \citep{Laughlin2004}. While the preference for hot Jupiters to orbit early M dwarfs may be an observational bias or a result of a small population size, the efficiency and prevalence of gas giants should increase when orbiting more massive M dwarfs.

\subsection{Comparison to planetary models}
These planets are unlikely to exhibit any radius inflation due to stellar flux-driven mechanisms. Studies of the \textit{Kepler} population of transiting hot Jupiters \citep[e.g.,][]{Thorngren2018,Thorngren2021} show that inflated radii are only evident in when $T_{eq}>1000$ K or the incident flux \(> 2 \times 10 ^{8}\mathrm{~erg~s^{-1}~cm^{-2}}\). These values serve as the threshold for the apparent radius anomaly of hot Jupiters, where some planets have radii much larger than expected for a pure hydrogen and helium Jovian analogue \citep[see][and references therein]{Fortney2021}. 

The mass and radii for TOI-3984 A and TOI-5293 A are within $2-3\sigma$ of the predicted values from models for gas giants between $1-5$ Gyrs by \cite{Baraffe2008}, which assume a gaseous hydrogen and helium envelope with a distribution of heavy elements, and \cite{Fortney2007}, which assume a solar metallicity hydrogen and helium atmosphere with a heavy element core composed of a $50/50$ mixture of ice and rock. We compared TOI-3984 A b and TOI-5293 A b to the predicted mass and radii for a solar metallicity atmosphere and note agreement in the mass and radius within $2-3\sigma$ regardless of age. For TOI-3984 A b, the models of \cite{Baraffe2008} suggest a large heavy element fraction ($Z=M_{\mathrm{env}}/M_p\approx0.3-0.5$) while the \cite{Fortney2007} models similarly suggest a core $0.25-0.55$ times the planetary mass. It may be possible the apparent upper limit for $Z$ is large because we ignore the effects of tidal heating. Through a population study of sub-Saturns, \cite{Millholland2020a} determined that ignoring radius inflation due to tides driven by a non-zero eccentricity or obliquity would result in ($Z\gtrsim0.5$). The younger models favor a larger $Z$, but the poor constraint on age and eccentricity preclude a detailed study on tidal heating in TOI-3984 A. In comparison, the mass and radius of TOI-5293 A are typical of a canonical Jupiter and are consistent with models from \cite{Fortney2007} having no core or the \cite{Baraffe2008} models with a negligible heavy element fraction of $Z=2\%$. 

\subsection{Future Characterization}
\subsubsection{Stellar Obliquity}
Measurements of the projected obliquity ($\lambda$), or the angle between a star's spin axis and the orbit normal of a companion, have been useful to constrain the physical processes responsible for the observed planetary architectures (see the reviews by \cite{Triaud2018} and \cite{Albrecht2022}). Studies of obliquities of hot Jupiters orbiting Sun-like stars have revealed an obliquity distribution that is consistent with a formation process involving high-eccentricity migration and tidal damping \cite[e.g.,][]{Albrecht2012,Rice2022}. 

Obliquity measurements for M dwarf systems are rarer due to the faintness of the host stars and only a handful of M dwarf systems have obliquity measurements: GJ 436 \citep{Bourrier2018,Bourrier2022}, K2-25 \citep{Stefansson2020b}, GJ 3740 \citep{Stefansson2022}, TRAPPIST-1 \citep{Hirano2020,Brady2023}, and AU Mic \citep{Hirano2020a,Palle2020,Addison2021}. Of these, GJ 436 b and GJ 3470 b are observed to be on polar orbits, while the other systems are observed to be on well-aligned orbits. Due to the rarity and relative faintness of M dwarfs hosting gas giants, there exists no measurement of $\lambda$ for such a system and it is not known if high-eccentricity migration coupled with tidal damping is also the dominant formation pathway for these planets.  

Both TOI-3984 A and TOI-5293 A have a measured rotation period (see Section \ref{sec:prot}), which allows for constraints on the three-dimensional obliquity, $\psi$ \cite[e.g.,][]{Stefansson2022,Frazier2023}. A measurement of $\psi$ for exoplanets across all spectral types has revealed these planets do not span the full range of $\psi$ but tend to be either polar or well-aligned, which may be a remnant of formation and not due to tidal evolution \citep{Albrecht2021,Spalding2022}. Measurements of $\psi$ for M dwarf gas giants are required to investigate if the distribution is also bimodal for this population. 

We estimate the amplitude of the RV anomaly due to the Rossiter-Mclaughlin effect \citep[Equation 1 in][]{Triaud2018} for both stars using the derived transit and stellar parameters. We derive an equatorial velocity of $v_{eq}=0.47\pm0.03~\mathrm{km~s^{-1}}$ and RM amplitude of \(\sim8~\mathrm{m~s^{-1}}\) for TOI-3984 A. For TOI-5293 A, $v_{eq}=1.28_{-0.04}^{+0.06}~\mathrm{km~s^{-1}}$ with an RM amplitude of \(\sim36~\mathrm{m~s^{-1}}\). The precision to detect these signals around faint ($V=16-17$) M dwarfs can be achieved using current high-resolution spectrometers on large telescopes, such as MAROON-X \citep{Seifahrt2016} or KPF \citep{Gibson2016}.

\subsubsection{Transmission Spectroscopy}
Understanding the atmospheric composition of an exoplanet is important to determine its bulk planetary composition and provide an constraints to internal structure models \citep[see the reviews by][]{Madhusudhan2019,Fortney2020} which can then be linked to formation and evolutionary processes. Theoretical studies have predicted that key volatile molecules, such as $\mathrm{H_2O}$, $\mathrm{CH_4}$, and $\mathrm{CO}$ should be present in the hydrogen-dominated atmospheres of hot Jupiters at high temperatures \citep[e.g.,][]{Burrows1999,Moses2011,Madhusudhan2012}. 
Current and future atmospheric missions, such as JWST \citep{Greene2016} and ARIEL \citep{Tinetti2018}, will characterize such gas giants over a wide range of temperatures to adequately sample a wide range of transitions in atmospheric chemistry \citep{Madhusudhan2011,Molliere2015,Fortney2020}. 

Beyond the detection of individual species, atmospheric abundances can be used to estimate the C/N/O ratios, which in turn are thought to trace where within the protoplanetary disk a planet forms   \citep{Oeberg2011,Madhusudhan2012,Oeberg2019,Turrini2021,Hobbs2022}. Current attempts to invert atmospheric composition to formation and migration processes have had limited success and more detailed characterization of exoplanet atmospheres would help inform planet formation models \citep[e.g.,][]{Dash2022,Molliere2022}.

\begin{figure*}[!ht]
\epsscale{1.15}
\plotone{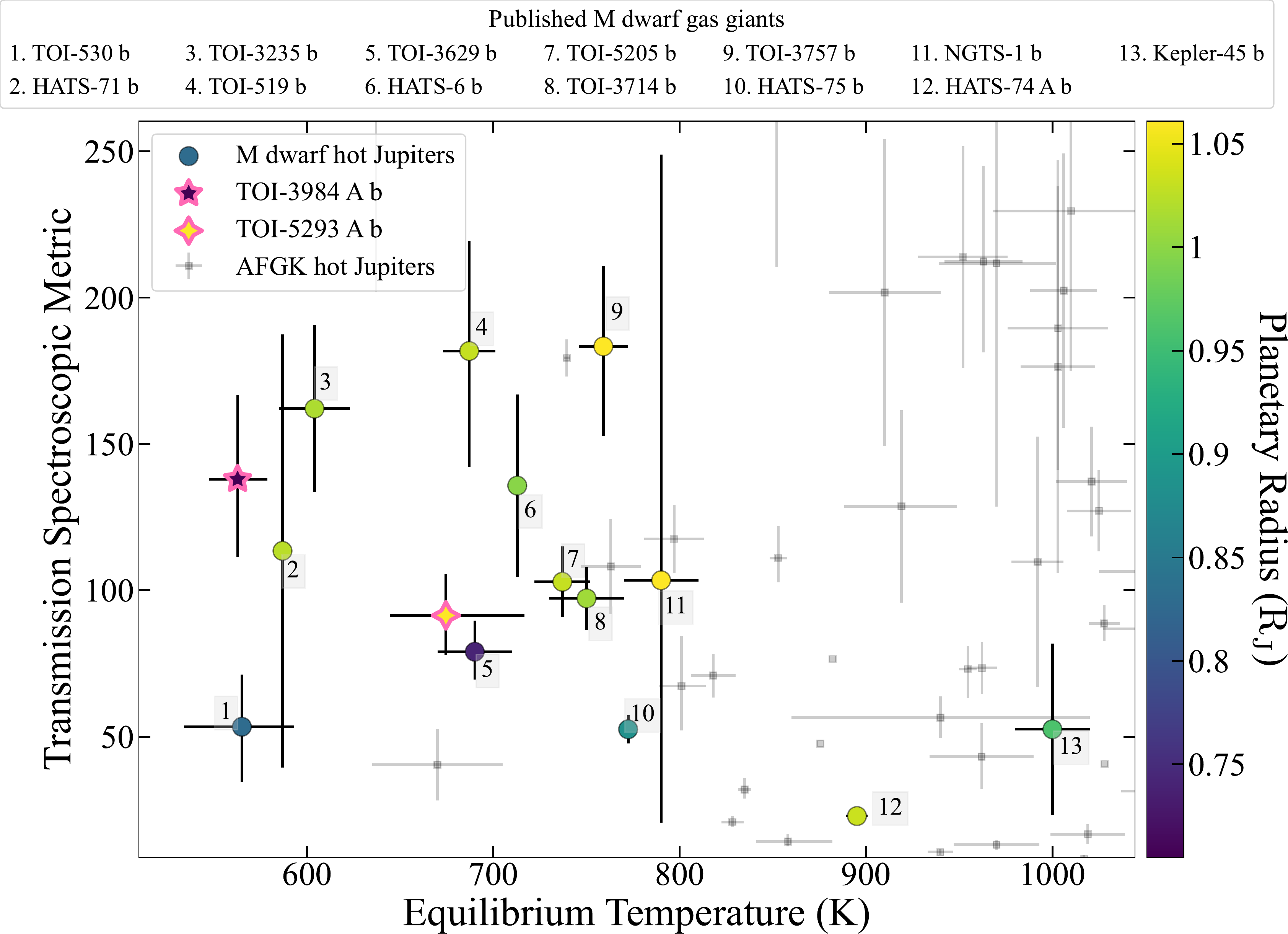}
\caption{A comparison of the transmission spectroscopic metric (TSM) from \cite{Kempton2018} for TOI-3984 A and TOI-5293 A to the existing M dwarf gas giant population ($P<10$ days and $R_p>8\mathrm{~R_J}$). TOI-3984 A is the coolest object in this sample and provides an opportunity to characterize the chemistry in a gas giant with $T_{eq}<600$ K. TOI-5293 A b overlaps with TOI-3629 b in equilibrium temperature and TSM but is much larger in radius and provides an opportunity to study a temperate gas giant with $R>1~\mathrm{R_J}$. The TSM value for hot Jupiters orbiting Sun-like stars are included for reference. The data were compiled from the \href{https://exoplanetarchive.ipac.caltech.edu/cgi-bin/TblView/nph-tblView?app=ExoTbls&config=PSCompPars}{NASA Exoplanet Archive} on 2023 May 03.}
\label{fig:tsm}
\end{figure*}

In the context of hot Jupiters orbiting Sun-like stars, studies with HST have revealed a diverse sample containing both cloudy and clear planets with depleted water abundance relative to predictions from the Solar System \cite[e.g.,][]{Sing2016,Pinhas2019,Welbanks2019}. No studies have been performed on M dwarf hot Jupiters to date. TOI-3984 A b and TOI-5293 A b have the precision on mass and radius ($>5\sigma$) required for detailed atmospheric analysis \citep{Batalha2019}. The large transmission spectroscopy metric \citep[TSM;][]{Kempton2018} of both TOI-3984 A b ($\mathrm{TSM}=138_{-27}^{+29}$) and TOI-5293 A b ($\mathrm{TSM}=92\pm14$) make these favorable targets among the M dwarf gas giant population to be observed with JWST, as shown in Figure \ref{fig:tsm}. While these planets do not have the highest TSM among the M dwarf gas giants (TOI-3757 has a $\mathrm{TSM}=180\pm30$), they are unique in the population due to their cooler equilibrium temperature. Almost all of the corresponding hot Jupiters orbiting Sun-like stars are hotter than TOI-3984 A b and TOI-5293 A b or have lower TSMs.    

An important factor in understanding exoplanetary atmospheres is the prevalence of clouds and hazes, which may form in the hot Jupiters through condensation chemistry or photochemical processes \citep[e.g.,][]{Sudarsky2003,Helling2008,Marley2013}. The presence of clouds or hazes can impact atmospheric processes and weaken or mask spectral features \citep[e.g.,][]{Sing2016,Sing2018}. Numerous theoretical studies have predicted the ubiquity of clouds in atmospheres at all temperatures \citep{Marley2015,Gao2021}. Disequilibrium processes are thought to be more efficient in the atmospheres cooler exoplanets like TOI-3984 A b and TOI-5293 A b \citep[e.g.,][]{Moses2011} and the production of aerosols may be enhanced due to the higher percentage of total UV energy from M dwarfs compared to Sun-like stars \citep[e.g.,][]{Liang2004,Line2010,Youngblood2016,Melbourne2020}. Transmission spectra with JWST using NIRSpec would provide ample wavelength coverage needed to make comparisons to predictions from cloud and haze models \citep[e.g.,][]{Kawashima2019,Mai2019} and the effects of higher UV radiation environment of early M dwarfs on atmospheric chemistry \citep[e.g.,][]{Pineda2021}.

\section{Summary}\label{sec:summary}
We present and characterize two gas giants orbiting mid M dwarfs. TOI-3984 A b is a short-period sub-Saturn ($M_{p}=0.14\pm0.03~\mathrm{M_J}$ and $R_{p}=0.71\pm0.02~\mathrm{R_J}$) on a $P=4.353326 \pm 0.000005$ day orbit. TOI-5293 A b is a hot Jupiter ($M_{p}=0.54\pm0.07~\mathrm{M_J}$ and $R_{p}=1.06\pm0.04~\mathrm{R_J}$) on a $P=2.930289 \pm 0.000004$ day orbit. Both systems have measured rotation periods between 10 and 70 days and probably ages between 0.71 and 5.1 Gyrs. They are in wide separation binary systems, with TOI-3984 A having a white dwarf companion at a projected separation of 356 au and TOI-5293 A having a later type M dwarf companion at a projected separation of 579 au. For TOI-3984 A, the age range estimated from the rotation period is in agreement with the nominal cooling age of its white dwarf companion ($\sim2.9$ Gyr), which may have once been close enough to impact the migration of TOI-3984 A b. The companion to TOI-5293 A is too far to impact the migration of its planet. Existing photometric and RV data are sufficient to reject the presence of additional massive, close-period planetary companions in these systems. TOI-3984 A b has the smallest mass and radius among the current sample of M dwarf gas giants while TOI-5293 A b is a typical M dwarf gas giant and consistent with the median mass and radius of the existing population. These two planets are, however, much cooler than the typical M dwarf gas giants due to the mid M dwarf nature of the host stars. TOI-3984 A b and TOI-5293 A b are bright enough to facilitate observations during transit to (i) further our understanding of their dynamical history with a measurement of the projected and three dimensional obliquities ($\lambda$ and $\psi$) and (ii) explore the atmospheric chemistry of temperate gas giants.

\section*{Acknowledgments}
We thank the anonymous referee for valuable feedback which has improved the quality of this manuscript. CIC acknowledges support by NASA Headquarters through an appointment to the NASA Postdoctoral Program at the Goddard Space Flight Center, administered by USRA through a contract with NASA and the NASA Earth and Space Science Fellowship Program through grant 80NSSC18K1114. 
SK acknowledges research support from Carnegie Institution of Science through the Carnegie Fellowship.
GS acknowledges support provided by NASA through the NASA Hubble Fellowship grant HST-HF2-51519.001-A awarded by STScI, which is operated by AURA, for NASA, under contract NAS5-26555.
The Center for Exoplanets and Habitable Worlds is supported by the Pennsylvania State University and the Eberly College of Science.

The computations for this research were performed on the Pennsylvania State University's Institute for Computational and Data Sciences' Roar supercomputer. This content is solely the responsibility of the authors and does not necessarily represent the views of the Institute for Computational and Data Sciences.

We acknowledge support from NSF grants AST 1006676, AST 1126413, AST 1310875, AST 1310885, AST 2009554, AST 2009889, AST 2108512, AST 2108801 and the NASA Astrobiology Institute (NNA09DA76A) in our pursuit of precision RVs in the near-infrared. We acknowledge support from the Heising-Simons Foundation via grant 2017-0494.

Some of these results are based on observations obtained with the Apache Point Observatory 3.5m telescope, which is owned and operated by the Astrophysical Research Consortium. 
We acknowledge support from NSF grants AST 1907622, AST 1909506, AST 1909682, AST 1910954 and the Research Corporation in connection with precision diffuser-assisted photometry.
We wish to thank the APO 3.5m telescope operators in their assistance in obtaining these data.

These results are based on observations obtained with HPF and LRS2 on the HET. The HET is a joint project of the University of Texas at Austin, the Pennsylvania State University, Ludwig-Maximilians-Universit\"at M\"unchen, and Georg-August Universit\"at Gottingen. The HET is named in honor of its principal benefactors, William P. Hobby and Robert E. Eberly. The HET collaboration acknowledges the support and resources from the Texas Advanced Computing Center. We are grateful to the HET Resident Astronomers and Telescope Operators for their valuable assistance in gathering our HPF data.
LRS2 was developed and funded by the University of Texas at Austin McDonald Observatory and Department of Astronomy and by Pennsylvania State University. We thank the Leibniz-Institut f\"ur Astrophysik Potsdam (AIP) and the Institut f\"ur Astrophysik G\"ottingen (IAG) for their contributions to the construction of the integral field units.
We would like to acknowledge that the HET is built on Indigenous land. Moreover, we would like to acknowledge and pay our respects to the Carrizo \& Comecrudo, Coahuiltecan, Caddo, Tonkawa, Comanche, Lipan Apache, Alabama-Coushatta, Kickapoo, Tigua Pueblo, and all the American Indian and Indigenous Peoples and communities who have been or have become a part of these lands and territories in Texas, here on Turtle Island.

Some of the data were obtained by the NEID spectrograph built by the Pennsylvania State University and operated at the WIYN Observatory by NOIRLab, which is managed by the Association of Universities for Research in Astronomy (AURA) under a cooperative agreement with the NSF, and operated under the NN-EXPLORE partnership of NASA and the NSF.
WIYN is a joint facility of the University of Wisconsin-Madison, Indiana University, NSF's NOIRLab, the Pennsylvania State University, Purdue University, University of California-Irvine, and the University of Missouri. 
Observations with NEID were obtained under proposal 2022A-763446 (PI: S. Wang). NEID results utilize the Data Reduction Pipeline operated by NExScI and developed under subcontract 1644767 between JPL and the University of Arizona. This work was performed for the Jet Propulsion Laboratory, California Institute of Technology, sponsored by the United States Government under the Prime Contract 80NM0018D0004 between Caltech and NASA. We thank the NEID Queue Observers and WIYN Observing Associates for their skillful execution of the NEID observations.
We thank Zade Arnold, Joe Davis, Michelle Edwards, John Ehret, Tina Juan, Brian Pisarek, Aaron Rowe, Fred Wortman, the Eastern Area Incident Management Team, and all of the firefighters and air support crew who fought the recent Contreras fire to save KPNO.
The authors are honored to be permitted to conduct astronomical research on Iolkam Du'ag (Kitt Peak), a mountain with particular significance to the Tohono O'odham.  

Some of the observations in this paper made use of the NN-EXPLORE Exoplanet and Stellar Speckle Imager (NESSI). NESSI was funded by the NASA Exoplanet Exploration Program and the NASA Ames Research Center. NESSI was built at the Ames Research Center by Steve B. Howell, Nic Scott, Elliott P. Horch, and Emmett Quigley.

This paper is based on observations obtained from the Las Campanas Remote Observatory through a partnership between Carnegie Observatories, the Astro-Physics Corporation, Howard Hedlund, Michael Long, Dave Jurasevich, and SSC Observatories.

Some of the data presented in this paper were obtained from MAST at STScI. The specific observations analyzed can be accessed via \dataset[DOI: 10.17909/1xcp-2f18]{https://doi.org/10.17909/1xcp-2f18}. Support for MAST for non-HST data is provided by the NASA Office of Space Science via grant NNX09AF08G and by other grants and contracts.
This work includes data collected by the TESS mission, which are publicly available from MAST. Funding for the TESS mission is provided by the NASA Science Mission directorate. 
This research made use of the (i) NASA Exoplanet Archive, which is operated by Caltech, under contract with NASA under the Exoplanet Exploration Program, (ii) SIMBAD database, operated at CDS, Strasbourg, France, (iii) NASA's Astrophysics Data System Bibliographic Services, (iv) NASA/IPAC Infrared Science Archive, which is funded by NASA and operated by the California Institute of Technology, and (v) data from 2MASS, a joint project of the University of Massachusetts and IPAC at Caltech, funded by NASA and the NSF.

This work has made use of data from the European Space Agency (ESA) mission Gaia (\url{https://www.cosmos.esa.int/gaia}), processed by the Gaia Data Processing and Analysis Consortium (DPAC, \url{https://www.cosmos.esa.int/web/gaia/dpac/consortium}). Funding for the DPAC has been provided by national institutions, in particular the institutions participating in the Gaia Multilateral Agreement.

Some of the observations in this paper made use of the Guoshoujing Telescope (LAMOST), a National Major Scientific Project built by the Chinese Academy of Sciences. Funding for the project has been provided by the National Development and Reform Commission. LAMOST is operated and managed by the National Astronomical Observatories, Chinese Academy of Sciences.

Some of the observations in this paper were obtained with the Samuel Oschin Telescope 48-inch and the 60-inch Telescope at the Palomar Observatory as part of the ZTF project. ZTF is supported by the NSF under Grant No. AST-2034437 and a collaboration including Caltech, IPAC, the Weizmann Institute for Science, the Oskar Klein Center at Stockholm University, the University of Maryland, Deutsches Elektronen-Synchrotron and Humboldt University, the TANGO Consortium of Taiwan, the University of Wisconsin at Milwaukee, Trinity College Dublin, Lawrence Livermore National Laboratories, and IN2P3, France. Operations are conducted by COO, IPAC, and UW.

\facilities{ARC (ARCTIC), Exoplanet Archive, Gaia, HET (HPF, LRS2), Las Campanas Observatory (TMMT, LCRO), LAMOST, MAST, PO:1.2m (ZTF), PO:1.5m (ZTF), TESS, WIYN (NESSI)} 
\software{
\texttt{astroquery} \citep{Ginsburg2019},
\texttt{astropy} \citep{AstropyCollaboration2018},
\texttt{barycorrpy} \citep{Kanodia2018}, 
\texttt{dynesty} \citep{Speagle2020},
\texttt{eleanor} \citep{Feinstein2019},
\texttt{EXOFASTv2} \citep{Eastman2019},
\texttt{GLS} \citep{Zechmeister2009},
\texttt{HPF-SpecMatch},
\texttt{juliet} \citep{Espinoza2019},
\texttt{lightkurve} \citep{LightkurveCollaboration2018},
\texttt{LRS2Multi},
\texttt{matplotlib} \citep{hunter2007},
\texttt{numpy} \citep{vanderwalt2011},
\texttt{Panacea},
\texttt{pandas} \citep{McKinney2010},
\texttt{PyHammer} \citep{Roulston2020},
\texttt{scipy} \citep{Virtanen2020},
\texttt{telfit} \citep{Gullikson2014},
\texttt{thejoker} \citep{Price-Whelan2017},
\texttt{TLS} \citep{Hippke2019},
\texttt{WD\_models} \citep{Cheng2019}
}

\appendix

\section{SED Fit to Broadband Photometry}\label{app:sed}
\begin{figure*}[!htb]
\epsscale{1.15}
\plotone{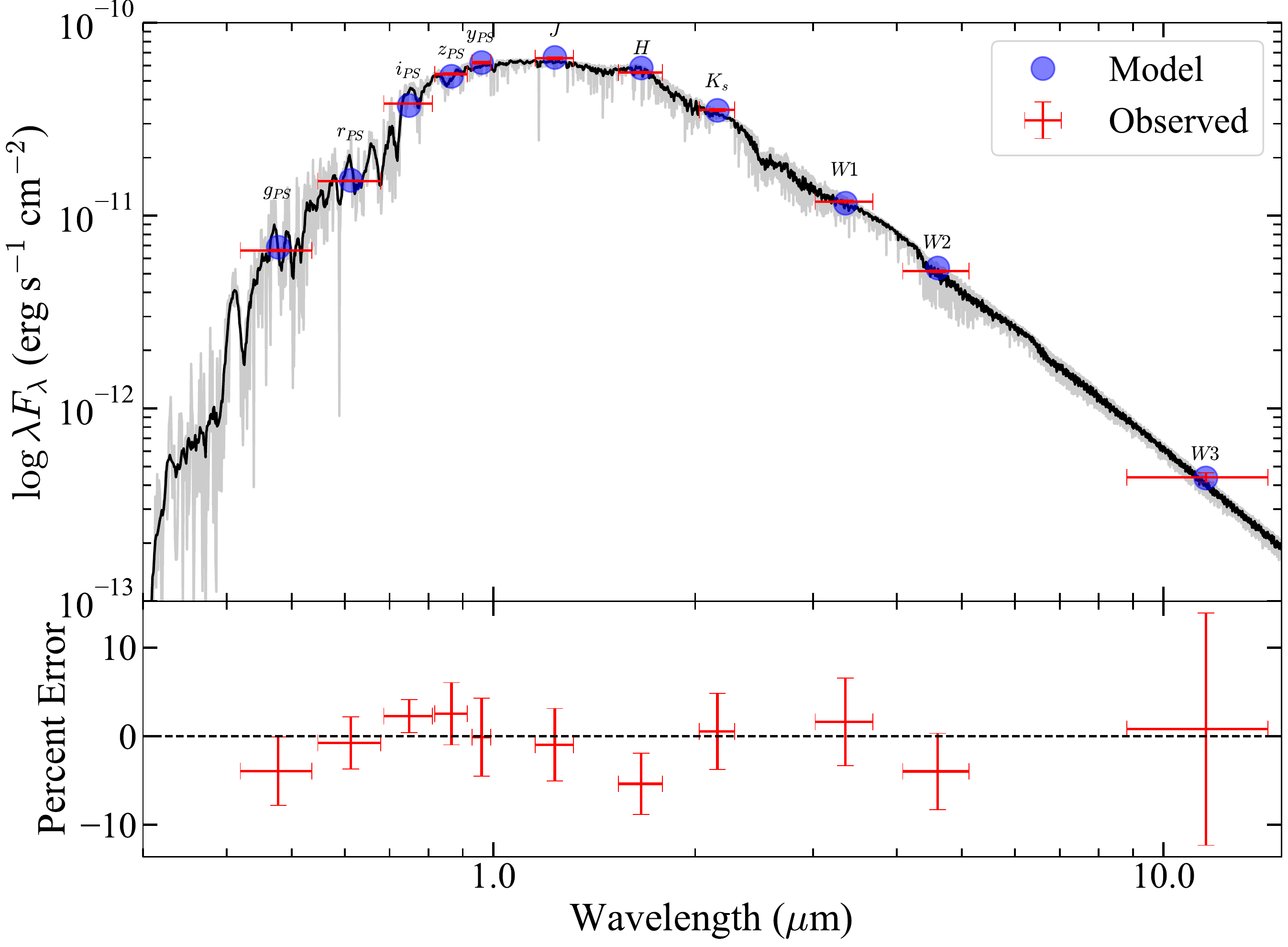}
\caption{The SED fit to TOI-3984 A. Red points are the measurements from broadband photometry while blue points represent the MIST model flux. The horizontal bars represent the width for each filter. A NextGen BT-SETTL model \citep{Allard2012} is overlaid for reference as a grey line and smoothed for clarity as a black line. The BT-SETTL model is not used as part of the SED fit.}
\label{fig:3984sed}
\end{figure*}

\begin{figure*}[!htb]
\epsscale{1.15}
\plotone{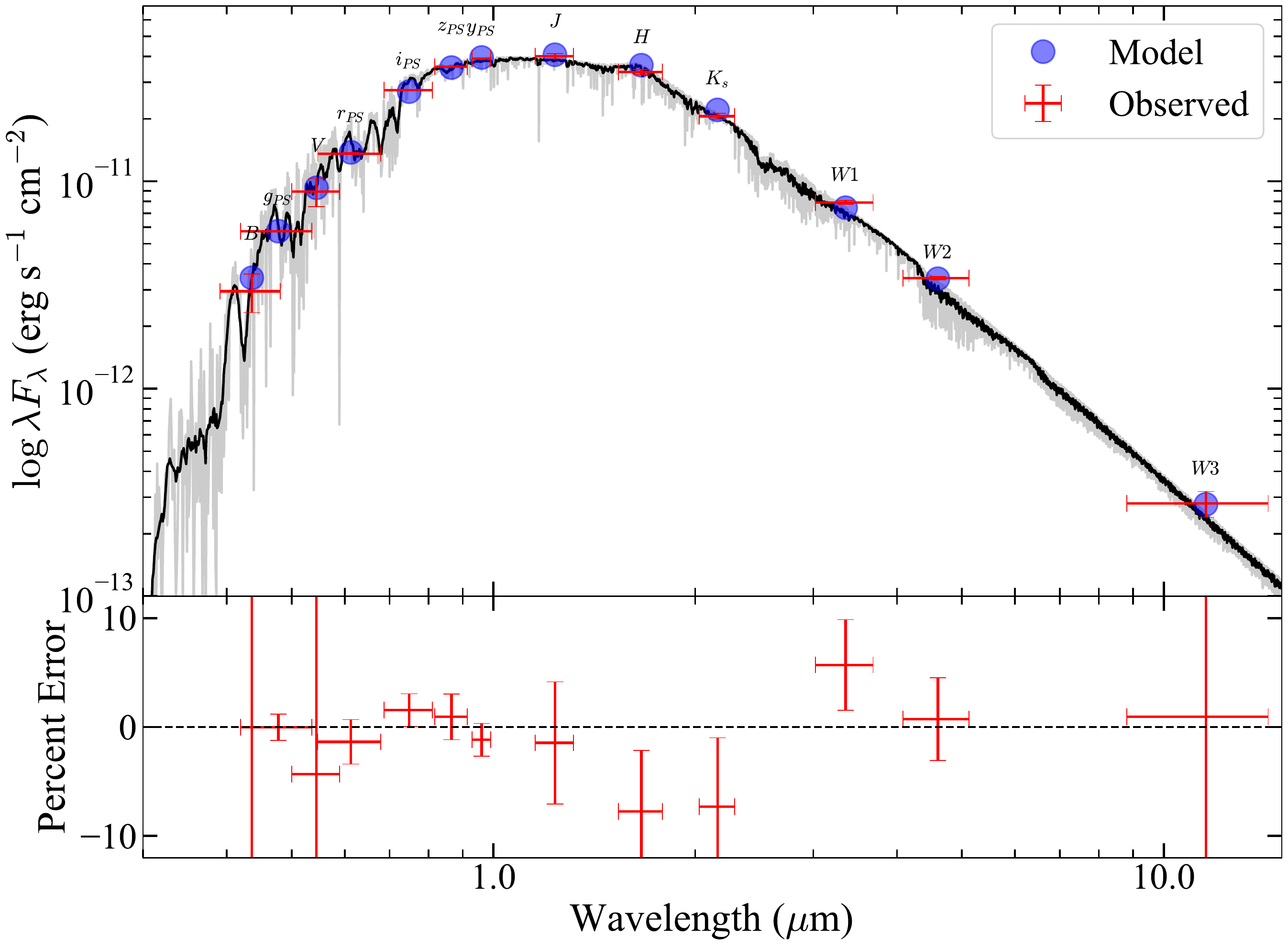}
\caption{Identical to Figure \ref{fig:3984sed} but for TOI-5293 A.}
\label{fig:5293sed}
\end{figure*}

The SED fits with \texttt{EXOFASTv2} use Gaussian priors on the (i) broadband photometry listed in Table \ref{tab:stellarparam}, (ii) $\log g_\star$, $T_{e}$, and [Fe/H] derived from \texttt{HPF-SpecMatch}, and (iii) the geometric distance calculated from \cite{Bailer-Jones2021} for each respective star. We apply an upper limit to the visual extinction based on estimates of Galactic dust \citep{Green2019} calculated at the distance determined by \cite{Bailer-Jones2021}. The \(R_{v}=3.1\) reddening law from \cite{Fitzpatrick1999} is used to convert the extinction from \cite{Green2019} to a visual magnitude extinction. The stellar parameters derived using the Gaia DR3 parallax are identical to the values derived with the \cite{Bailer-Jones2021} distance priors. This is expected because \cite{Bailer-Jones2021} note that the inverse parallax provides a good distance estimate for stars with positive parallaxes and a ratio between the parallax error to the parallax of $\sigma_\varpi/\varpi<0.1$. Figures \ref{fig:3984sed} and \ref{fig:5293sed} present the SED fits for TOI-3984 A and TOI-5293 A, respectively. 

\section{Measuring the rotation period}\label{app:prot}
\subsection{TESS}
We do not search for photometric modulation using the TESS PDCSAP flux or \texttt{eleanor} \texttt{CORR\_FLUX} because the algorithms that produce these light curves are known to attenuate and distort long-period ($>10$ day) astrophysical signals, such as starspot-induced photometric variability \citep[see][]{Gilliland2015,VanCleve2016,Feinstein2019,Holcomb2022}. We use the TESS Systematics-insensitive Periodogram package \citep[\texttt{TESS-SIP};][]{Hedges2020} to simultaneously detrend systematics in the uncorrected flux and create a Lomb-Scargle periodogram. \texttt{TESS-SIP} uses a linear model consisting of regressors (for this search, 3 principal components and a mean offset) and a sinusoid component. We limit the search to periods between $1-60$ days and analyzed all available long-cadence data after excising the transits and bad quality data (non-zero TESS quality flags) for each star. No significant period for TOI-3984 A and TOI-5293 A were recovered. 
\subsection{ZTF}
TOI-3984 A and TOI-5293 A were observed with ZTF as part of a survey of the TESS northern sectors \citep{vanRoestel2019} and the photometry is publicly available under DR15\footnote{\url{https://www.ztf.caltech.edu/ztf-public-releases.html}}. ZTF has a plate scale of $1.012\arcsec{}\mathrm{~pixel^{-1}}$ \citep{Yao2019} and all exposures for these objects are 30s. We use the constraints from the ZTF Science Data System Explanatory Supplement\footnote{\url{https://web.ipac.caltech.edu/staff/fmasci/ztf/ztf_pipelines_deliverables.pdf}} (ZDS) and reject observations with (i) non-zero \texttt{catflag} values (see \textsection 13.6 in ZDS), (ii) values of $\chi<0.5$ and $\chi>1.5$, where $\chi$ is the rms of the residuals to the PSF fit on the source performed by the ZTF pipeline, (iii) values of $|\mathtt{sharp}|\ge0.5$, where \texttt{sharp} is the difference of the observed and model squared PSF FWHM, and (iv) airmass $>1.8$. TOI-3984 A has (i) 471 observations in the $zg$ filter between 2018 March 25 and 2022 July 22 with a median cadence and precision of 1 day and $\sim1.3\%$ and (ii) 498 observations in the $zr$ filter between 2018 March 21 and 2022 July 17 with a median cadence and precision of 1 day and $\sim1.1\%$, respectively. TOI-5293 A has (i) 197 observations in the $zg$ filter between 2018 June 26 and 2022 November 06 with a median cadence and precision of 3 days and $\sim1.5\%$ and (ii) 181 observations in the $zr$ filter between 2018 June 29 and 2022 October 26 with a median cadence and precision of 3 days and $\sim1.1\%$, respectively. 

To derive the rotation period, we modeled the ZTF photometry with the \texttt{juliet} analysis package \citep{Espinoza2019} which performs the parameter estimation using the dynamic nested-sampling algorithm \texttt{dynesty} \citep{Speagle2020}. The photometric model consists of a Gaussian process noise model with the approximate quasi-periodic covariance function from \cite{Foreman-Mackey2017}. This kernel has allowed for computationally efficient inference of stellar rotation periods in large datasets that are not uniformly sampled \citep[e.g.,][]{Angus2018}.

\begin{figure*}[!htb]
\epsscale{1.05}
\plotone{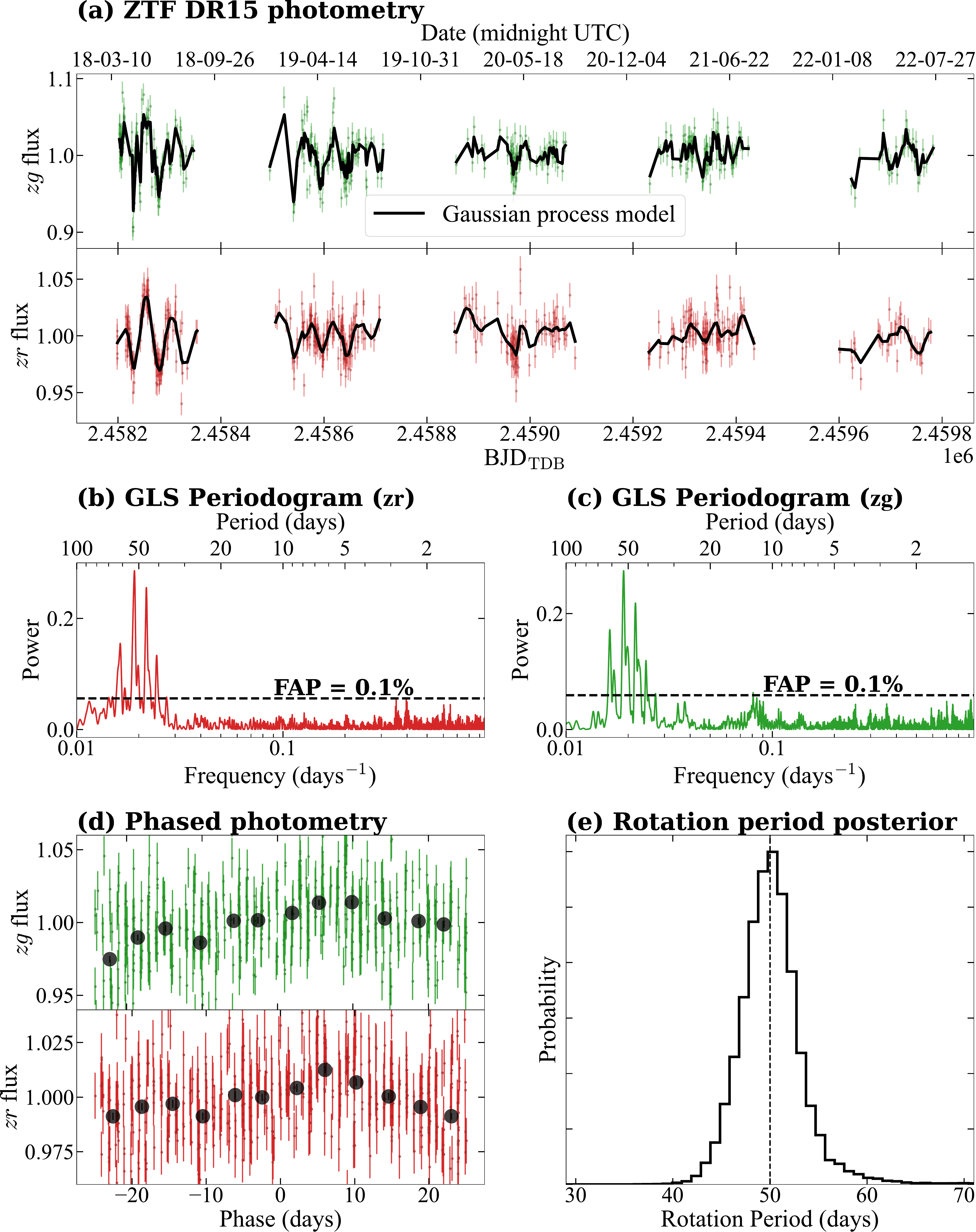}
\caption{\textbf{(a)} displays the ZTF photometry for TOI-3984 A in each filter along with the best-fitting Gaussian process model for reference. \textbf{(b)}  and \textbf{(c)} are the GLS periodograms for the $zr$ and $zg$ photometry from ZTF. A false-alarm probability of 0.01\% \citep[following][]{Zechmeister2009} is shown for reference. \textbf{(d)} presents the ground-based ZTF photometry from panel (a) phased to the derived rotation period. The large black points represent 4-day bins of the phased photometry. \textbf{(e)} presents the posterior distribution of the rotation period from the Gaussian process model. The derived rotation period is \(50.0_{-2.7}^{+2.8}\) days.}
\label{fig:3984phasedrot}
\end{figure*}

\begin{figure*}[!htb]
\epsscale{1.1}
\plotone{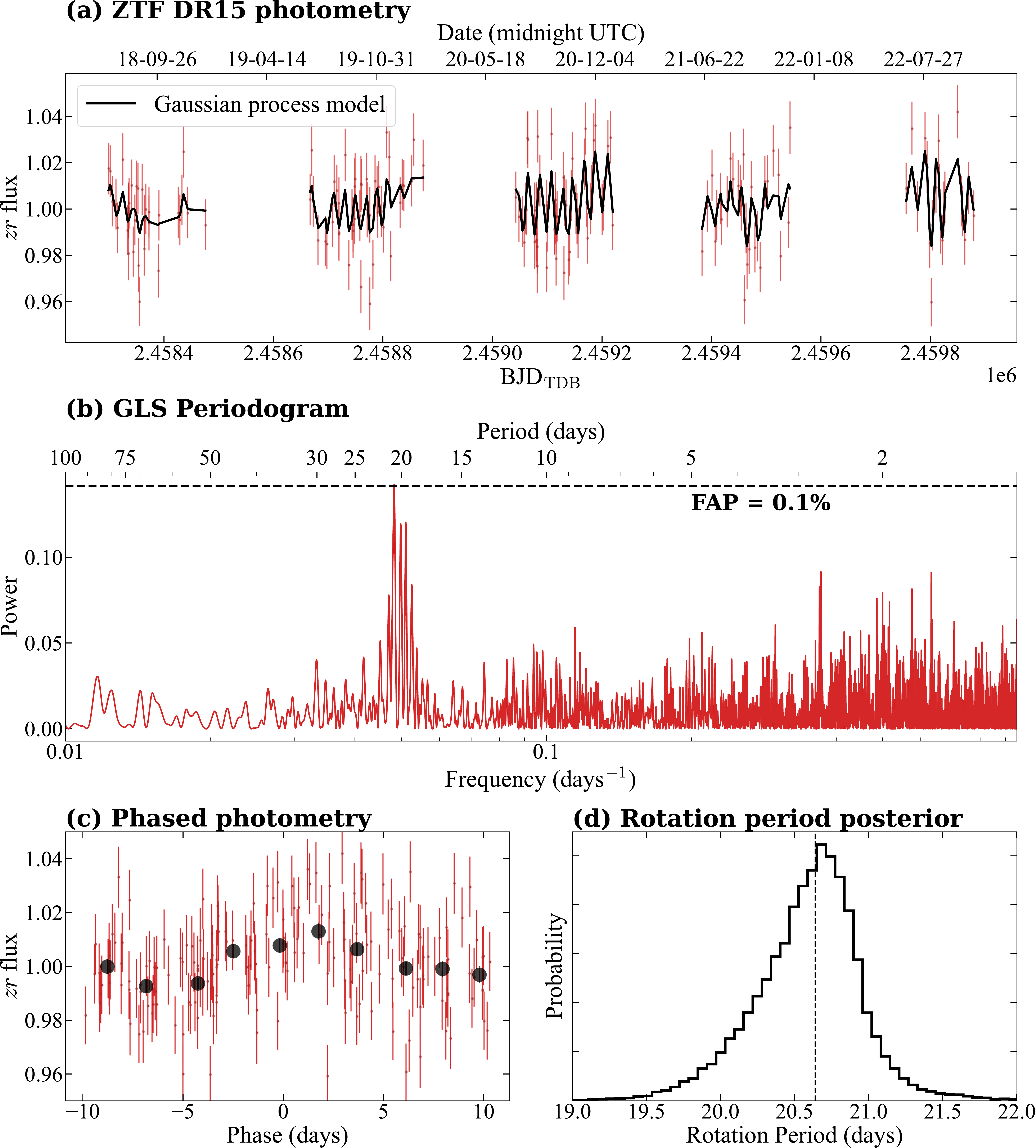}
\caption{Identical to Figure \ref{fig:3984phasedrot} but for TOI-5293 A. \textbf{(a)} displays the ZTF photometry along with the best-fitting Gaussian process model for reference. \textbf{(b)} presents the GLS periodogram for the ZTF data. \textbf{(c)} presents the phased ground-based ZTF photometry with large black points representing 2-day bins. \textbf{(d)} presents the posterior distribution of the rotation period. The derived rotation period is \(20.6_{-0.4}^{+0.3}\) days.}
\label{fig:5293phasedrot}
\end{figure*}

We interpret the periodicity of the Gaussian process kernel as the stellar rotation period and force the same kernel periodicity for both ZTF filters of the same star. We include a simple white-noise model $\sigma_{\mathrm{phot}}$ in the form of a jitter term that is added in quadrature to the error bars of the photometry for each filter. The fit for each star uses a uniform prior on the Gaussian process period of $1.1-100$ days, where the upper limit reflects the upper limits from other photometric surveys of mid M dwarfs \citep{McQuillan2013,Newton2018}. The priors and posterior values for the fits to ZTF photometry are listed in Table \ref{tab:ztfpriors}. The ZTF data and best fitting Gaussian process kernel are presented in Figure \ref{fig:3984phasedrot} and \ref{fig:5293phasedrot}.

\begin{deluxetable*}{llccc}
\tabletypesize{\footnotesize}
\tablecaption{ZTF photometric modeling \label{tab:ztfpriors}}
\tablehead{\colhead{~~~Parameter} &
\colhead{Units} &
\colhead{Prior} &
\colhead{TOI-3984 A} &
\colhead{TOI-5293 A}
}
\startdata
\sidehead{$zg$ Gaussian process parameters:}
~~~$B$ & Amplitude (ppm) & $\mathcal{J}(10^{-4},10^{12})$ & $470_{-60}^{+70}$ & \nodata \\
~~~$C$ & Additive factor  & $\mathcal{J}(10^{-3},10^3)$ &  $0.8_{-0.8}^{+100}$ & \nodata \\
~~~$L$ & Length scale (days)  & $\mathcal{J}(10^{-3},10^3)$ &  $4.6_{-0.9}^{+1.2}$ & \nodata  \\
~~~Photometric jitter & $\sigma_{phot}$ (ppm) & $\mathcal{J}(10^{-6},10^{3})$ & $0.03_{-0.03}^{+45.07}$ & \nodata \\
\hline
\sidehead{$zr$ Gaussian process parameters:}
~~~$B$ & Amplitude (ppm) & $\mathcal{J}(10^{-4},10^{12})$ & $200_{-40}^{+60}$ & $160_{-50}^{+60}$ \\
~~~$C$ & Additive factor  & $\mathcal{J}(10^{-3},10^3)$ &  $0.02_{-0.02}^{+0.21}$ & $0.03_{-0.02}^{+0.3}$ \\
~~~$L$ & Length scale (days)  & $\mathcal{J}(10^{-3},10^3)$ &  $60_{-20}^{+40}$ & $120_{-70}^{+200}$  \\
~~~Photometric jitter & $\sigma_{phot}$ (ppm) & $\mathcal{J}(10^{-6},10^{3})$ & $0.03_{-0.03}^{+43.36}$ & $0.05_{-0.05}^{+42.17}$ \\
\hline
\sidehead{Shared Gaussian process parameters:}
~~~$P_{GP}$ & Period (days)  & $\mathcal{J}(1.1,100)$ &  $50.0_{-2.7}^{+2.8}$ & $20.6_{-0.4}^{+0.3}$ \\
\hline
\enddata
\end{deluxetable*}

\section{White dwarf companion to TOI-3984 A}\label{app:wdloc}
\begin{figure*}[!ht]
    \epsscale{1.15} 
    \plotone{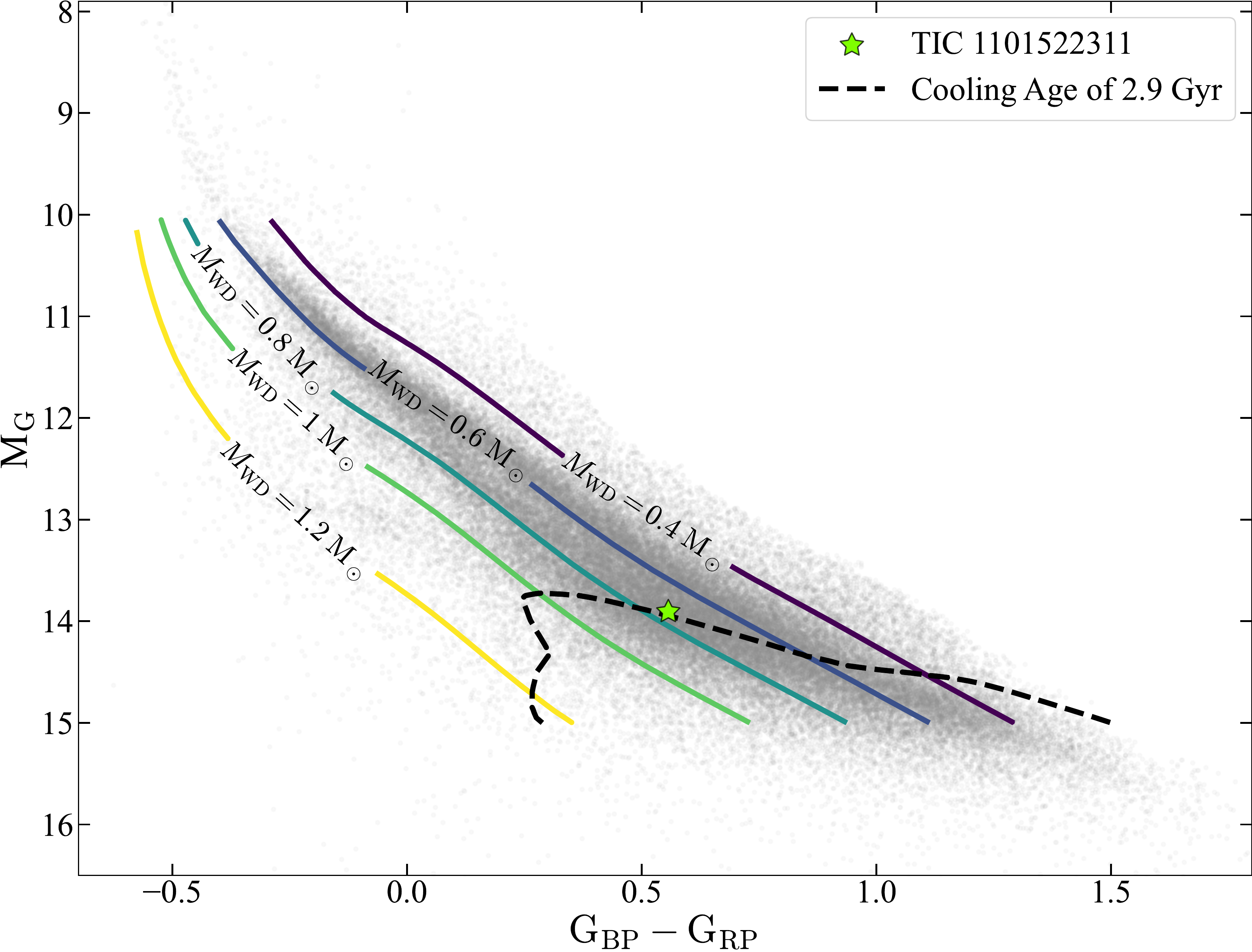}
    \caption{The nominal position of TIC 1101522311, the white dwarf companion to TOI-3984 A, on the color-magnitude diagram for white dwarfs identified in Gaia EDR3 by \cite{GentileFusillo2021}. Contours for fixed masses from \cite{Bedard2020} are plotted for reference. The best-matching cooling track from the models is shown with a dashed line.} \label{fig:wdloc}
\end{figure*}

We use the nominal Gaia DR3 colors to predict the mass and cooling age using the \texttt{WD\_models} package and the models from \cite{Bedard2020} for a hydrogen dominated atmosphere. Figure \ref{fig:wdloc} displays the location of TOI-3984 B (TIC 1101522311) on the Gaia color-magnitude diagram \citep[data from][]{GentileFusillo2021} along with the best-matching model used to derive a photometric age and mass from its location.

\bibliography{combined}
\bibliographystyle{aasjournal}



\end{CJK*}
\end{document}